\documentclass[usenatbib]{mn2e}
\usepackage{apjfonts}
\usepackage{epsfig}
\usepackage{amssymb}
\usepackage{amsmath}
\usepackage{fixltx2e} 
\usepackage{url}

\newcommand{\mergercalcurl}{\url{http://www.cfa.harvard.edu/~phopkins/Site/mergercalc.html}}
\newcommand\plotone[1]
 {\centering \leavevmode \includegraphics[width={0.99\columnwidth}]{#1}}
\newcommand\plotonesmall[1]
 {\centering \leavevmode \includegraphics[width={0.95\columnwidth}]{#1}}
\newcommand{\plotside}[1]
 {\centering \leavevmode \includegraphics[width={0.95\textwidth}]{#1}}
\newcommand{\plotsidesmall}[1]
 {\centering \leavevmode \includegraphics[width={0.85\textwidth}]{#1}}
\newcommand{\acknowledgments}{\begin{small}\section*{Acknowledgments}\end{small}}
\newcommand\altaffilmark[1]{$^{#1}$}
\newcommand\altaffiltext[1]{$^{#1}$}
\newcommand{\etal}{et al.}

\newcommand{\mstar}{M_{\ast}}
\newcommand{\msun}{M_{\sun}}

\title[Testing Models for Spheroid Size Evolution]{Discriminating Between the 
Physical Processes that Drive Spheroid Size Evolution}

\author[Hopkins \etal]{
\parbox[t]{\textwidth}{ 
Philip F.\ Hopkins\altaffilmark{1}\thanks{E-mail:phopkins@astro.berkeley.edu},
Kevin Bundy\altaffilmark{1},
Lars Hernquist\altaffilmark{2},
Stijn Wuyts\altaffilmark{2,3},
\&\ Thomas J.\ Cox\altaffilmark{2,3}, 
} 
\vspace*{6pt} \\
\altaffiltext{1}{Department of Astronomy, University of California 
Berkeley, Berkeley, CA 94720} \\
\altaffiltext{2}{Harvard-Smithsonian Center for Astrophysics, 60
  Garden Street, Cambridge, MA 02138, USA} \\ 
\altaffiltext{3}{W.~M.\ Keck Postdoctoral Fellow} 
}

\date{Submitted to MNRAS, August 10, 2009}
\begin{document}
\maketitle
\label{firstpage}

\begin{abstract}

Massive galaxies at high-$z$ have smaller effective radii than those today, but
similar central densities. Their size growth therefore relates primarily to the
evolving abundance of low-density material. Various models have been proposed to
explain this evolution, which have different implications for galaxy, star, and
BH formation. We compile observations of spheroid properties as a function of
redshift and use them to test proposed models. Evolution in progenitor
gas-richness with redshift gives rise to initial formation of smaller spheroids
at high-$z$. These systems can then evolve in apparent or physical size via
several channels: (1) equal-density `dry' mergers, (2) later major or minor
`dry' mergers with less-dense galaxies, (3) adiabatic expansion, (4) evolution
in stellar populations \&\ mass-to-light-ratio gradients, (5) age-dependent bias
in stellar mass estimators, (6) observational fitting/selection effects. If any
one of these is tuned to explain observed size evolution, they make distinct
predictions for evolution in other galaxy properties. Only model (2) is
consistent with observations as a dominant effect. It is the only model which
allows for an increase in $M_{\rm BH}/M_{\rm bulge}$ with redshift. Still, the amount of
merging needed is larger than that observed or predicted. We therefore compare
cosmologically motivated simulations, in which all these effects occur, \&\ show
they are consistent with all the observational constraints. Effect (2), which
builds up an extended low-density envelope, does dominate the evolution, but
effects 1, 3, 4, \&\ 6 each contribute $\sim20\%$ to the size evolution (a net factor $\sim2$).
This naturally also predicts evolution in $M_{\rm BH}-\sigma$ similar to that observed.

\end{abstract}


\section{Introduction}
\label{sec:intro}

Observations have suggested that 
high-redshift spheroids have significantly smaller effective 
radii than low-redshift analogues of the same mass 
\citep[e.g.][]{daddi05:drgs,vandokkum:z2.sizes,
zirm:drg.sizes,trujillo:size.evolution,
franx:size.evol,damjanov:red.nuggets,
vanderwel:z1.compact.ell,cimatti:highz.compact.gal.smgs,
trujillo:ell.size.evol.update,trujillo:compact.most.massive,toft:z2.sizes.vs.sfr,
buitrago:highz.size.evol}. Whatever 
process explains this apparent evolution must be particular to 
this class of galaxies: disk galaxies do not become similarly compact 
at high redshift \citep[][and references therein]{somerville:disk.size.evol,
buitrago:highz.size.evol}.\footnote{
There are of course different 
ways of defining galaxy ``density'' or ``compactness,'' 
for which the disk/spheroid difference is not the same 
\citep[see e.g.][]{buitrago:highz.size.evol}. 
In this paper, we will generally use these terms to refer to 
the coarse-grained phase-space density, 
which can be approximated by 
$f \sim M / (R^{3}\,V^{3}) \sim 1/(G\,R^{2}\,V)$, 
where $R$ is the characteristic major-axis radius 
(proportional to e.g.\ effective radius of spheroids or disk scale-length), 
and $V$ the characteristic velocity (dispersion or circular 
velocity). This is the quantity of theoretical interest as, 
in dissipationless processes, it is conserved or decreased 
\citep[$f_{\rm final} \le f_{\rm initial}$; 
see e.g.\ the discussion in][]{hernquist:phasespace}. Quantities such as 
the orbital streaming motion and disk scale heights 
enter in the fine-grained phase-space density, which although 
formally conserved is not observable. Similarly, as discussed in 
\S~\ref{sec:fgas.evol}, we take ``size'' to refer to the 
semi-major axis half-light size. 
}
In addition, these high-redshift 
populations have been linked to observed sub-millimeter galaxies, 
the most rapidly star-forming objects in the Universe, and 
bright, high-redshift quasar hosts \citep{younger:smg.sizes,
tacconi:smg.mgr.lifetime.to.quiescent,hopkins:groups.qso,
hopkins:groups.ell,alexander:smg.bh.masses}. 
As such, these observations represent a strong 
constraint on models of galaxy and bulge formation. 

More recently, \citet{hopkins:density.galcores} showed that the 
{\em central} densities of these high-redshift systems are similar 
to those of massive ellipticals at low redshifts; the primary difference 
between ``small'' (high-redshift) and ``large'' (low-redshift) systems 
relates to the amount of observed low-density material at 
large radii, absent in the high-redshift systems 
\citep[see also][]{bezanson:massive.gal.cores.evol}. There 
are therefore two important, related questions. First, 
how do high-redshift massive spheroids form, apparently without 
low density material, but with their dense cores 
more or less in place relative to their $z=0$ descendants? 
And second, do these early-forming systems ``catch up'' 
to later-forming massive counterparts by accumulating such 
low-density material? If so, how?

High-resolution hydrodynamic simulations have shown, for 
example, that the size of a spheroid {\em at the time of 
its formation} primarily reflects the 
degree of dissipation involved -- i.e.\ the loss of angular 
momentum by disk gas and its participation in a dense, central 
starburst \citep{cox:kinematics,
onorbe:diss.fp.details,ciotti:dry.vs.wet.mergers,
jesseit:kinematics,hopkins:cusps.ell,hopkins:cores}. 
If all the mass of a spheroid were formed in such a starburst, 
then one would expect an extremely small size $\lesssim$\,kpc, 
comparable to the sizes of e.g.\ ULIRG starburst regions. If, on the other 
hand, none of the mass were formed in this way, the size of the remnant 
would simply reflect the (large, $\sim5-10\,$kpc) extents of 
disk/star-forming progenitors. 
This leads to a natural expectation for 
size evolution. 

Disks at $z=2$ are observed to be much more 
gas-rich than those of comparable mass today 
\citep[e.g.][]{erb:lbg.gasmasses}, so their mergers will naturally 
lead to smaller remnants \citep{khochfar:size.evolution.model,
hopkins:bhfp.theory,hopkins:cusps.evol}. 
In fact, simulations have shown that simply scaling extended, low-density 
disks in the local Universe to the typical gas fractions of $\sim L_{\ast}$ 
disks at $z>2$ is sufficient to produce $\sim 10^{11}\,\msun$ 
ellipticals with $R_{e}\sim1\,$kpc \citep{hopkins:cusps.fp}. 
Ellipticals that form at lower redshift (most of the population), 
from less gas-rich mergers, will be larger, so the mean size-mass 
relation will evolve. In particular, these systems, forming more of their 
mass from violent relaxation of the (low-density) progenitor stellar disks (those 
being relatively gas-poor, yielding little starburst), will have a 
larger ``envelope'' of low density material at large radii. 

Meanwhile, the compact, early-forming ellipticals will undergo 
later ``dry'' mergers, in particular with later-forming, more gas-poor disks 
and less dense ellipticals (less dense because they formed later, from 
less gas-rich disks, as described above). This will preferentially add 
mass to their low-density profile ``wings'' 
and increase $R_{e}$, allowing them to ``catch up'' to later forming systems.  
Thus not only does such a general scenario anticipate evolution in the 
median size-mass evolution, but also the nature of such evolution: 
early buildup of dense regions via dissipation, followed by later 
growth in the extended, low-density wings as the progenitor population becomes 
less gas-rich with cosmic time. 

That being said, cosmological galaxy formation models have difficulty explaining 
how systems could catch up from 
the most extreme size evolution seen in the observations: a 
factor of $\sim6$ smaller {\em effective} radii $R_{e}$ 
at fixed stellar mass in the most 
massive galaxies at $z=2$ \citep[see][]{toft:z2.sizes.vs.sfr,
buitrago:highz.size.evol}. The models above predict 
a more moderate factor $\sim3$ evolution 
in the most massive systems. In addition, 
the expected efficiency of mergers should leave some 
galaxies un-merged (at least without major re-mergers or 
several minor re-mergers) since their original gas-rich, 
spheroid-forming merger \citep{hopkins:cusps.evol}. If 
no other effects act on these systems, this surviving 
fraction would overpredict 
the number of such compact systems today \citep{trujillo:dense.gal.nearby,
taylor:2009.no.compact.massive.local.gal}
\citep[but see also][]{valentinuzzi:superdense.local.ell.wings}.
Moreover, some local observations have argued that, at fixed stellar mass, 
ellipticals with older stellar population ages and/or those in the 
most dense (and early-forming) environments may have the 
{\em largest} radii for their mass -- if so, these systems 
have evolved to ``overshoot'' the median of the local size-mass relations, 
evolving by more like a factor $\sim10$ in effective radius 
\citep{gallazzi06:ages,bernardi:bcg.scalings,graves:ssps.vs.fp.location}. 
(Although this depends on whether ``mass'' is defined by dynamical 
or stellar mass, and we note \citealt{vanderwel:size.numden.massive.ell,
valentinuzzi:superdense.local.ell.wings}, who reach different conclusions.) 
As a consequence, other interpretations of the observations and 
theoretical models for size evolution have been debated. 

In this paper, we show how dissipation drives the formation of 
smaller ellipticals at high redshift (\S~\ref{sec:fgas.evol}) and 
consider the different explanations that have 
been proposed for how these massive, high-redshift ellipticals increase 
their apparent sizes at lower redshifts, 
and construct the predictions made by each individual model 
for other observable quantities, including their velocity dispersions, 
central densities, masses, and profile shapes 
(\S~\ref{sec:models}). We compile observations of these 
quantities and other constraints to break the degeneracies between 
the different models (\S~\ref{sec:problems}). We then compare a model 
motivated by cosmological simulations, in which many of 
these effects occur at different points in the 
galaxy's evolution. We show how such a mixed model tracks through 
the predicted space, and how relatively small contributions from 
each of the proposed explanations combine to yield 
order-of-magnitude cumulative size evolution (\S~\ref{sec:mixed}). 
We summarize and discuss our conclusions in \S~\ref{sec:discuss}. 

Throughout, we assume a WMAP5 \citep{komatsu:wmap5} cosmology, 
and a \citet{chabrier:imf} stellar IMF, 
but the exact choices make no significant difference to our conclusions.

\section{Why Does the Size-Mass Relation Evolve in the First Place?}
\label{sec:fgas.evol}

\begin{figure}
    %\plotone{evol_absence.ps}
    \plotone{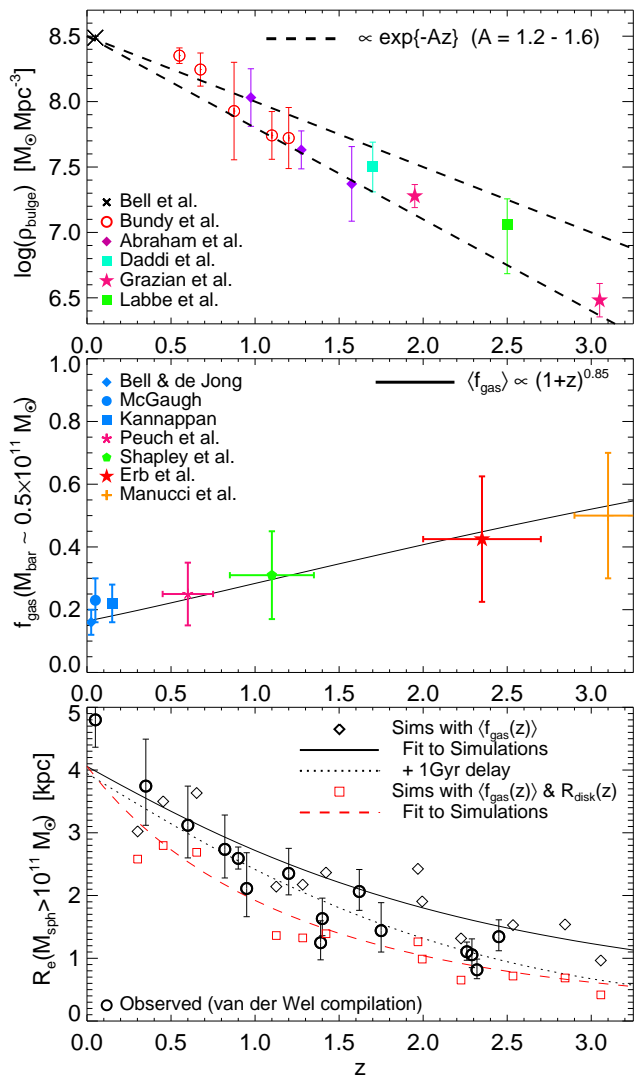}
    \caption{{\em Top:} 
    Evolution in the mass density in spheroid-dominated 
    (compact) galaxies. Points are observations; dashed lines a fit. 
    Evolution is steep; at all $z$, {\em most} spheroids are recently-formed 
    and have {\em not} experienced e.g.\ dry mergers. 
    The size-mass relation must primarily reflect how 
    spheroids form {\em in situ}. 
    {\em Middle:} 
    Evolution in typical gas fractions of star-forming (spheroid progenitor) 
    galaxies of total (baryonic) mass such that their major merger will 
    yield a $\gtrsim10^{11}\,\msun$ spheroid. Points observed; error bars 
    show the scatter in $f_{\rm gas}$ at fixed mass, not the (smaller) uncertainty in the mean. 
    Line shows a fit. 
    {\em Bottom:} 
    Predicted sizes of $\sim10^{11}\,\msun$ ellipticals formed {\em in situ} 
    at each redshift from gas-rich mergers with the expected $\langle f_{\rm gas}(z) \rangle$. 
    Circles show observed (mean) sizes compiled in \citet{vanderwel:z1.compact.ell}. 
    Black diamonds show actual hydrodynamic simulation 
    remnants with the same progenitor disk size/structure but with different gas 
    fractions appropriate for each $\langle f_{\rm gas}(z) \rangle$. 
    Black solid line shows a fit, given the median scalings of $\langle f_{\rm gas}(z) \rangle$ 
    and $R_{e}(f_{\rm gas})$. 
    Dotted line is the same, allowing for a $1\,$Gyr delay after each merger before 
    it is observed as a ``passive'' remnant. 
    Red diamonds are similar simulations, but also include 
    viewing biases and stellar population effects (e.g.\ mock images at the appropriate times) and 
    allow for the maximum observationally inferred (weak) progenitor disk size evolution 
    ($R_{\rm disk}\propto (1+z)^{0.3-0.5}$). Red dashed line shows the appropriate 
    fit including this additional scaling.  
    Evolving gas-richness and rapid new buildup of ellipticals drives
    the evolution in the size-mass relation. Dry mergers and other effects do not 
    dominate the relation, but explain how 
    early-forming systems ``catch up to'' (or exceed) the relation at low redshifts. 
    \label{fig:evol.nodry}}
\end{figure}

First, we must consider how small, high-mass ellipticals are formed initially.
This is discussed in detail in e.g.\ \citet{khochfar:size.evolution.model,
naab:size.evol.from.minor.mergers,feldmann:bgg.size.evol.in.hydro.sims} and \citet{hopkins:cusps.evol}; 
but we briefly review these results here. 

Figure~\ref{fig:evol.nodry} summarizes the important physics. 
First, consider the mass density in passive ellipticals as 
a function of redshift \citep[{\em top}; from][]{bell:mfs,
bundy:mfs,bundy:mtrans,abraham:red.mass.density,daddi05:drgs,
labbe05:drgs,vandokkum06:drgs,grazian:drg.comparisons}.\footnote{Specifically,
  we plot the mass density in bulge-dominated galaxies,
  which is not the same as the absolute mass density in all bulges. 
  At high redshifts 
  $z>1.5$ observed morphologies are ambiguous; we show 
  the mass density in passively
  evolving red galaxies as a proxy. This may not be appropriate, 
  but at $z<1$ the two correspond well, and 
  the compactness, size, and kinematics of the ``passive'' objects 
  do appear distinct from star-forming ones 
  \citep{kriek:drg.seds,toft:z2.sizes.vs.sfr,trujillo:ell.size.evol.update,
  franx:size.evol,genzel:highz.rapid.secular}.
  }
It is well-established that this declines rapidly with redshift; 
we fit the observations shown in Figure~\ref{fig:evol.nodry} 
with the simple functional form 
$\rho_{\rm ell}\propto \exp{(-A\,z)}$ and find $A\approx1.2-1.6$. 
In other words, at $z=2$, only $\sim5\%$ of the $z=0$, $\sim L_{\ast}$ elliptical 
mass density is in place. It is possible to consider this in greater detail 
in terms of number counts or as a function of galaxy mass, 
but the qualitative results are similar in each 
case: the population ``in place'', represented by the compact, high-redshift 
massive elliptical population, is only a small fraction of the population that 
will be present at any significantly lower redshift 
\citep[e.g.][]{vandokkum:z2.sizes,kriek:highz.red.sequence,
perezgonzalez:mf.compilation,marchesini:highz.stellar.mfs,
vanderwel:size.numden.massive.ell,ilbert:cosmos.morph.mfs}.\footnote{
Although there is some debate regarding the degree of evolution in 
number density of the very most massive galaxies at 
$z<0.8$, the $\sim L_{\ast}$, $M_{\ast}\sim10^{11}\,\msun$ population 
on which we focus here (and which defines the observed samples 
to which we compare) dominates the effects shown in Figure~\ref{fig:evol.nodry} 
and clearly shows a rapid decrease with redshift even over this 
range \citep[see e.g.\ the references above and][]{bundy:mfs,
pannella:mfs,franceschini:mfs,borch:mfs,brown:mf.evolution,
pozzetti:2009.z1.gal.bimodal.mf}. Moreover, 
the rapid decline in passive spheroid number density with redshift is 
clear at all masses at $z\gtrsim1.5$, the range of 
particular interest here. 
}
As such, evolution in the {\em median} size-mass 
relation -- i.e.\ the average size of galaxies at a given stellar mass -- must 
reflect evolution in the sizes {\em at the time of formation}. 

This is very important: at any redshift, {\em most} of the spheroid population 
is {\em recently formed}, and has {\em not} had to evolve from some earlier 
redshift via e.g.\ dry mergers or any other channel. 
The evolution in the size-mass relation cannot, therefore, be the result of 
all ellipticals forming early (with some size) and then dry merging or experiencing 
other processes that increase their size to $z=0$. Rather, at each redshift, 
the size of ellipticals forming at that time must be larger than the size 
at the time of formation of ellipticals that formed earlier. These ``newly formed'' 
spheroids, which comprise most of the population, will dominate the size-mass 
relation at that redshift, independent of how the earlier-forming population evolves 
(which constitute part of the scatter, but do not dominate the median relation). 

What, therefore, determines the sizes of ellipticals at formation? 
At both low and high redshifts, star-forming disk and/or rotationally supported 
galaxies have much larger sizes than ellipticals of 
similar mass \citep[see e.g.][]{kormendy:spheroidal1,
shen:size.mass,trujillo:size.mass.to.z3,toft:z2.sizes.vs.sfr,
buitrago:highz.size.evol}. 
Moreover the {\em slope} of the disk/rotationally supported galaxy 
size-mass relation is distinct from that of the spheroid size-mass 
relation at all $z=0-3$ (references above). And, in addition, observations 
have shown that disk sizes do not appear to evolve 
with redshift nearly as strongly as 
spheroid sizes \citep{trujillo:size.mass.to.z3,
ravindranath:disk.size.evol,ferguson:disk.size.evol,
barden:disk.size.evol,toft:z2.sizes.vs.sfr,akiyama:lbg.weak.size.evol,
buitrago:highz.size.evol,somerville:disk.size.evol}. 
Parameterizing size evolution with redshift (at fixed mass) 
as a power-law $R_{e}(M_{\ast}\,|\,z)\propto (1+z)^{-\beta}$, 
the observations constrain the {\em maximum} $\beta$ for disk 
galaxies to be $\beta < 0.6$ at 
intermediate masses and $\beta<0.8$ at the highest masses 
(with many of the observations still consistent 
with $\beta\approx0$ for disks). Massive 
ellipticals, on the other hand, appear to evolve with $\beta\approx1.4-1.7$.  
It is clear, therefore, that the sizes of ellipticals at formation, and the evolution in 
their size-mass relation, cannot simply reflect the sizes of their 
progenitors. 

Simple phase-space considerations, however, make it impossible to increase 
densities in dissipationless (purely stellar) mergers \citep{hernquist:phasespace}. 
But in sufficiently gas-rich mergers, the remnant size can be much smaller 
than that of the progenitor. Gas dissipation makes 
this possible; disk gas loses angular momentum via internal torques in the 
merger \citep{barnes.hernquist.91,barneshernquist92}, and can then dissipate energy, 
fall to the center, and build stars in a compact central starburst on scales 
$\ll$\,kpc \citep{mihos:starbursts.94,mihos:starbursts.96}. 
High-resolution hydrodynamic simulations, and basic 
physical arguments, have shown that this degree of dissipation 
is the primary determinant of the remnant spheroid size 
\citep{cox:kinematics,robertson:fp,
naab:gas,onorbe:diss.fp.details,ciotti:dry.vs.wet.mergers,
jesseit:kinematics,jesseit:merger.rem.spin.vs.gas,
covington:diss.size.expectation,
hopkins:cusps.ell,hopkins:cores}. 
To lowest order, this degree of dissipation -- i.e.\ the mass fraction formed 
dissipationally -- simply reflects the 
cold gas fractions available in the progenitor disks at the time of the 
merger \citep{hopkins:disk.survival}. 
To rough approximation, one can fit 
the results of high-resolution simulations of gas-rich mergers (references 
above) and estimate how the remnant spheroid size scales with this 
gas fraction: 
$R_{e}(M_{\ast}\,|\,f_{\rm gas})\approx R_{e}(M_{\ast}\,|\,f_{\rm gas}=0)\,\exp{(-f_{\rm gas}/0.3)}$. 
(Based on the arguments above, $R_{e}(M_{\ast}\,|\,f_{\rm gas}=0)$ is 
equivalent, modulo a geometric prefactor, to the pre-merger disk effective 
radii.) 
If the gas mass is sufficiently large ($\sim1/2$ the galaxy mass), then the remnant 
size (half-mass radius) will simply reflect the sub-kpc scales of the 
starburst region; if it is sufficiently small, it has no effect and the remnant sizes 
reflect those of disk progenitors. 

The natural expectation, of course, is that the gas fractions of 
higher-redshift star-forming galaxies will be larger than those of 
low-redshift systems of the same mass; 
this has now been seen in a number of direct observations. 
Figure~\ref{fig:evol.nodry} ({\em middle}) compiles observations 
from \citet{belldejong:tf}, \citet{kannappan:gfs}, and 
\citet{mcgaugh:tf} at low redshift, and 
\citet{shapley:z1.abundances,erb:lbg.gasmasses,
puech:tf.evol,mannucci:z3.gal.gfs.tf} at redshifts $z\sim1-3$ 
\citep[see also][]{calura:sdss.gas.fracs,
forsterschreiber:z2.sf.gal.spectroscopy,erb:outflow.inflow.masses}.\footnote{At $z=0$, 
  the gas fractions shown are based
  on measured atomic HI gas fractions; \citet{belldejong:tf} correct
  this to include both He and molecular H$_{2}$; \citet{mcgaugh:tf}
  correct for He but not H$_{2}$; \citet{kannappan:gfs} gives just the
  atomic HI gas fractions \citep[this leads to slightly lower
    estimates, but still within the range of uncertainty plotted;
    H$_{2}$ may account for $\sim20-30\%$ of the dynamical mass, per
    the measurements in][]{jogee:H2.masses}.  We emphasize that these
  gas fractions are lower limits (based on observed HI flux in some
  annulus).  At $z=2$,
  direct measurements are not always available; the gas masses from
  \citet{erb:lbg.gasmasses} are estimated indirectly based on the
  observed surface densities of star formation and assuming that the
  $z=0$ Kennicutt law holds; other indirect estimates 
  yield similar results \citep{cresci:dynamics.highz.disks}.} 
Specifically, we plot the gas fractions 
observationally inferred for disk/rotationally dominated, 
star-forming galaxies of {\em baryonic} masses $\sim0.5\times10^{11}\,\msun$ 
(the stellar masses may be a factor of a couple lower, corresponding 
to the observed $f_{\rm gas}$) 
such that after a major merger (which will increase the total mass 
and turn the gas into stars), the system will be comparable to the 
$\sim10^{11}\,\msun$ observed massive, compact ellipticals.
Figure~\ref{fig:evol.nodry} shows that $f_{\rm gas}$ 
grows from $\sim0.15-0.20$ at $z=0$ to $\sim0.4-0.5$ 
at $z=2.5-3$ (approximately as $\langle f_{\rm gas} \rangle = 0.16\,(1+z)^{0.85}$). 

This leads to an expected evolution in the sizes of 
spheroids at their time of formation. A detailed set of predictions for 
this size evolution as a result of evolving gas fractions is presented 
in \citet{hopkins:bhfp.theory,hopkins:cusps.evol} and a similar 
model in \citet{khochfar:size.evolution.model}; in Figure~\ref{fig:evol.nodry} 
we simply summarize the key result. 
Figure~\ref{fig:evol.nodry} ({\em bottom}) plots the expected 
size of major gas-rich $\sim 10^{11}\,\msun$ merger 
remnants, considering the mean gas fractions as a function of redshift (shown above). 
We compare these with the observed average sizes of spheroids of the given 
mass at each redshift. 
We show the results of simulated remnants from \citet{hopkins:cusps.ell}, 
with the same initial disk sizes appropriate for this mass, but with the 
relevant $f_{\rm gas}$ for the median at each redshift. 
We also show the corresponding median trend estimated by simply 
combining the fitted $R_{e}(M_{\ast}\,|\,f_{\rm gas})$ scaling above 
with the $\langle f_{\rm gas}(z) \rangle$ scaling. 
In these cases the disks are all just as extended as low-redshift 
disks (i.e.\ yield $R_{e}(10^{11}\,\msun\,|\,f_{\rm gas}=0) \approx 6-7$\,kpc, 
with no evolution in disk size with redshift). Already, 
this appears sufficient to explain at least the high end of the observed 
evolution. 
We could also consider the trend if we allow for a delay between merger and 
observation (because the observed systems are passive, one might 
preferentially select systems that had their gas-rich merger $\sim$Gyr ago; 
and so had correspondingly somewhat higher $f_{\rm gas}$ reflecting the 
expectation at that earlier time); this gives similar but slightly stronger evolution. 
We can also allow for some moderate disk size evolution. 
If we scale the sizes of progenitor disks (and, as a consequence, 
the remnant $R_{e}$ in the absence of gas) by 
the maximum allowed by the observations, $\propto (1+z)^{-0.5}$, 
we again find similar but slightly stronger evolution. 
These simple size predictions agree very well with the observed 
spheroid sizes at all intermediate and high redshifts. 

In other words, {\em it is straightforward to form $\sim1\,$kpc-sized 
$\sim 10^{11}\,\msun$ ellipticals at $z=2$, and in fact such sizes 
are the natural expectation given the observed/expected gas-richness of 
spheroid-forming mergers at these redshifts.} 
The difficulty is {\em not} ``how to form'' such ellipticals, nor is it even to 
explain the average evolution in the size-mass relation. 
Rather, the difficulty is that, as discussed in \S~\ref{sec:intro}, 
such systems clearly do not passively evolve to $z=0$ (where they 
would constitute a small but easily detectable fraction of the 
local spheroid population); in fact there does 
not even appear to be evidence for a small fraction of such systems 
retaining their small sizes to $z=0$. It may even 
be the case that early forming systems 
would to be the {\em largest} for their mass (i.e.\ have lower 
{\em effective} densities) at $z=0$ 
\citep[see e.g.][and references in 
\S~\ref{sec:intro}]{graves:ssps.vs.fp.location}. Although this 
remains observationally unclear, it is generally agreed that 
the most highly-clustered systems, massive BGGs and BCGs (the 
expected descendants of the most massive, 
first-to-assemble high-redshift systems) appear to lie 
significantly {\em above} the 
\citet{shen:size.mass} size-mass relation for more recently-assembled 
field galaxies \citep[see e.g.][]{batcheldor:bcgs,vonderlinden:bcg.scaling.relations,
jk:profiles,lauer:massive.bhs,bernardi:bcg.scalings}. 

Some process, therefore, not only increases the sizes of these high-redshift 
early-forming systems so as to ``keep pace'' with the mean evolution in 
the size-mass relation, but may even need to ``overshoot'' the relation. 
Some hint of this can be seen even in Figure~\ref{fig:evol.nodry}; 
at the lowest redshifts, assuming all ellipticals are formed in situ 
from gas-rich mergers actually {\em under}-predicts the median low-redshift 
sizes. By this late time (unlike at high redshifts) a non-trivial fraction of the 
population has formed earlier and undergone some subsequent evolution, 
bringing {\em up} the average size at these masses. 
In what follows, therefore, we consider how such systems might 
grow in size at a pace equal to or greater than the rate of change in the 
size of newly-forming systems shown in Figure~\ref{fig:evol.nodry}. 

Note that the sizes shown in Figure~\ref{fig:evol.nodry}, and the simulation sizes 
to which we refer throughout this paper (and, where possible, the observed sizes) 
refer to the {\em semi-major axis} lengths, for 
elliptical systems. This is almost identical to the (projected) circular radius $R$ 
that encloses $1/2$ the light (on average, the two are the same, in simulations). 
However, this is {\em not} identical to the ``circularized'' radius 
$R_{\rm circ} \equiv \sqrt{a\,b}$, where $a$ and $b$ are the major and 
minor axis lengths ($a\approx R_{e}$), respectively ($\epsilon\equiv1-b/a$ being the 
standard definition of ellipticity). 
The reason for our choice is that the major axis length or projected half-light circular radius 
is more physically robust, and relevant to the constraints from phase space densities 
and merger histories. A thin disk, for example, viewed edge-on, has a vanishingly 
small circularized radius (arbitrarily small $b$), even though the parameter of 
physical interest, the scale length, is the same. Examination of the suite of simulations 
shown here, for example, shows that while at low ellipticity systems can be either large 
or small, there are no systems with very large ellipticity and large $R_{\rm circ}$, 
even though such systems are merely flattened relative to their counterparts, and have similar 
scale lengths, velocity dispersions, energetics, and physical phase space densities. 
Following \citet{dehnen:1993.hernquist.profile.families}, 
one can show analytically that dissipationlessly randomizing a system 
which is flattened owing to some rotation, conserving total energy and phase space density, 
leads to very little change in the projected major-axis radius (this is the basic reason why the projected 
$R_{e}$ of dissipationless disk-disk merger remnants is similar to the in-plane $R_{e}=1.65\,h$ 
of their progenitor disks, discussed above), but will obviously increase $R_{\rm circ}$ by 
an arbitrarily large factor depending on the thickness of the original system. 
Because more gas-rich mergers lead to remnants with more rotation (on average), 
the ellipticity of the systems in Figure~\ref{fig:evol.nodry} can be somewhat higher at high 
redshift, and therefore their circularized radii evolve even more steeply 
(averaging over all inclinations, though, the effect is weak, adding a power 
$\sim(1+z)^{-0.25}$ to the redshift evolution).

\section{The Models and Their Predictions}
\label{sec:models}

\begin{figure*}
    %\plotside{demo_size_evol_models.ps}
    \plotside{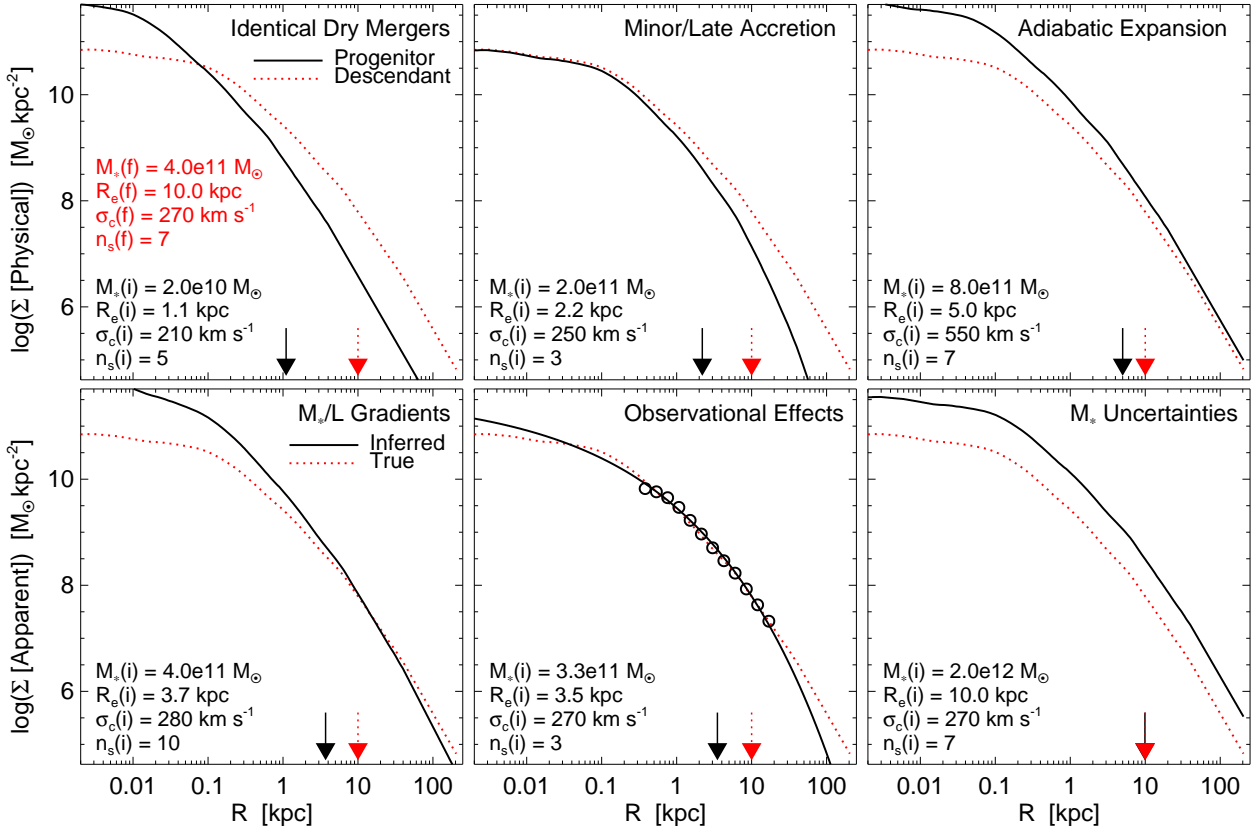}
    \caption{Evolution of physical ({\em top}) or observationally 
    inferred ({\em bottom}) surface stellar mass density profiles, according to different models. 
    We consider six models, described in the text: the physical 
    stellar mass profile can change owing to identical dry mergers 
    (doubling $M_{\ast}$ and $R_{e}$; {\em top left}), 
    minor/late accretion (building up an extended envelope; {\em top center}), or 
    adiabatic expansion (mass loss leading to uniform inflation; {\em top right}). 
    The inferred mass profile (fitted from observations, assuming standard 
    stellar populations and constant $M_{\ast}/L$)
    can change owing to 
    the presence of stellar mass-to-light ratio gradients (from young 
    central stellar populations; {\em bottom left}), 
    seeing effects and surface brightness dimming (points show 
    a simulated $z=2$ profile with typical seeing and surface brightness depth, 
    solid line is the best-fit $r^{1/4}$-law profile given the observed range; {\em bottom center}), 
    or discrepancies between the true and best-fit 
    stellar mass (owing to e.g.\ contribution of AGB stars; {\em bottom right}). 
    In each case, the plotted initial profile is a factor of $\sim2-3$ ``too small'' relative to the 
    $z=0$ size-mass relation. Each model is tuned so that, after evolution according to the 
    model, the same final profile (in this case that matching the observed profile of a 
    typical massive core elliptical on the local size-mass relation, NGC 4365) is recovered. 
    Properties of the initial galaxies and final remnant (same in all cases) are given; 
    arrows show effective radii. 
    \label{fig:ex.sbevol}}
\end{figure*}

We consider the following sources of size evolution for systems 
initially formed at some high redshift ($z\sim2$) as compact, massive 
galaxies ($R_{e}\sim1\,$kpc, $M_{\ast}\sim10^{11}\,\msun$), 
illustrated in Figure~\ref{fig:ex.sbevol}: \\ 

{\bf ``Identical'' Dry Mergers:} Often, when the term ``dry mergers'' is used in the 
context of models for size evolution, what is actually assumed is not general 
gas-poor merging but specifically spheroid-spheroid re-mergers, between spheroids 
with otherwise identical properties (or at least identical profile shapes and 
effective densities, in the case of non-equal mass mergers). In a 1:1 such merger, 
energetic arguments as well as simulations \citep{hernquist:phasespace,
hopkins:cores} imply profile 
shape and velocity dispersion are conserved, while mass and size double 
(more generally, $R_{e}\propto M$). 
\\

{\bf Minor/Late Accretion:} In fact, the scenario above is {\em not} expected to be a primary 
growth channel for massive ellipticals, nor is it expected to be the most common form 
of ``dry merger.'' Rather, in models,  
at later times the typical secondary (even in 
major mergers) is a later-forming galaxy -- a gas-poor disk or more ``puffy'' 
spheroid that was itself formed from more gas-poor mergers and therefore less compact 
\citep[][]{hopkins:cusps.evol,naab:size.evol.from.minor.mergers,feldmann:bgg.size.evol.in.hydro.sims}.
At the highest masses, growth preferentially becomes dominated by more and more minor 
mergers with low-effective density galaxies \citep{maller:sph.merger.rates,hopkins:merger.rates}. 
The secondaries, being lower-density, build up extended ``wings'' around the high central 
density peak in the elliptical. The central density and velocity dispersion remain 
nearly constant, while the Sersic index of the galaxy increases with the buildup of these 
wings \citep[see e.g.][]{naab:profiles,hopkins:cores,hopkins:disk.survival}. 
The effective radius grows much faster per unit mass added in these mergers -- 
it is possible to increase $R_{e}$ of a given elliptical by a factor $\sim6$ per mass doubling 
($R_{e}\propto M^{2-2.5}$). 
\\

{\bf Adiabatic Expansion:} If a system loses mass from its central regions in an adiabatic 
manner, the generic response of stars and dark matter will be to ``puff up,'' as the central 
potential is less deep. For a spherically symmetric, 
homologous contraction of shells with circular orbits, this reduces to the 
criterion that $M(r)\,r=$\,constant. More general scenarios behave in a 
similar manner \citep[modulo small corrections; see][]{
zhao:2002.adiabatic.expansion.stellarwinds,
gnedin:adiabatic.contraction}. 
If a galaxy could lose a large fraction of its central 
(baryon-dominated) mass, either by efficiently expelling the material from 
stellar mass loss or by blowing out a large fraction of the baryonic 
mass in an initially very gas-rich system 
\citep[from e.g.\ quasar feedback;][]{fan:adiabatic.expansion.ell.size.evol}, the 
radius will grow correspondingly ($R_{e}\propto M^{-1}$). The profile 
shape will be conserved (to lowest order, although this 
depends at second order on the stellar age distribution versus radius)
but uniformly inflated, the central density will 
decrease sharply, and the velocity dispersion will decrease $\propto R_{e}^{-1}$. 
\\

{\bf $M_{\ast}/L$ Gradients:} Massive, old ellipticals at $z=0$ have weak color 
gradients \citep[][and references 
therein]{mcdermid:sauron.profiles,sanchezblazquez:ssp.gradients}. 
As such, the effective radius in certain optical bands is generally a good proxy for the 
stellar mass $R_{e}$. However, the same is not necessarily true for young 
ellipticals recently formed in mergers; these can have blue cores at their 
centers with young stars just formed in the merger-driven starburst 
\citep[e.g.][]{rj:profiles}. Simulations and resolved stellar population analysis 
suggests that the resulting gradient in $M_{\ast}/L$ (brighter towards the center) 
can lead to smaller $R_{e}$ by up to a factor $\sim2$ in optical bands (e.g.\ rest-frame 
$B$), relative to the stellar mass $R_{e}$ \citep{hopkins:cusps.mergers}. 
As the system ages, these gradients will vanish and $R_{e}$ in optical bands 
will appear to increase; moreover, depending on the exact band observed, the late-time 
$R_{e,\,\rm light}$ may actually over-estimate the stellar mass $R_{e}$ (as e.g.\ age gradients 
fade and long-lived metallicity gradients remain, yielding a redder center and hence 
apparently less concentrated optical light distribution). Thus at both early and late times, 
$M_{\ast}/L$ gradients can, in principle, yield evolution in the size at fixed wavelength, 
while conserving the stellar mass $R_{e}$. 
Obviously, the central mass density will remain 
constant, and the velocity dispersion will only weakly shift (with 
the appropriate luminosity-weighting). 
\\

{\bf Seeing/Observational Effects:} The large effective 
radii of massive, low-redshift ellipticals are driven by material in 
low surface density ``envelopes'' at large radii. This is difficult to recover 
at high redshifts. Moreover, galaxy surface 
brightness profiles are not perfect \citet{sersic:profile} profiles, so the 
best-fit Sersic profile and corresponding $R_{e}$ will depend on the 
dynamic range observed \citep[see e.g.][]{boylankolchin:mergers.fp,
hopkins:density.galcores}. Together, these effects can, in principle, lead to a smaller 
{\em fitted} $R_{e}$ (from sampling only the central regions, inferring a 
smaller $n_{s}$ and less low-density material) at high redshifts 
(although it is by no means clear that the biases {\em must} go in this direction). 
At lower redshifts, where surface brightness limits are less severe, 
more such material would be recovered, leading to apparent size-mass 
and profile shape evolution ($R_{e}$ changes at fixed $M_{\ast}$), 
without central surface density or velocity dispersion evolution. 

Another effect that might occur is that high-redshift ellipticals could be 
more flattened than those at low redshift. If the definition of effective 
radius used is that of the ``circularized'' radius $R_{\rm circ}=\sqrt{a\,b}$, then 
a flattened high-redshift system could have small $R_{\rm circ}$ from 
edge-on sightlines (and smaller median $R_{\rm circ}$, by a lesser factor), 
per the discussion above in \S~\ref{sec:fgas.evol}.  
Indeed, many such compact systems are observed to be relatively 
elliptical \citep{valentinuzzi:superdense.local.ell.wings,vandokkum:z2.sizes}. 
This could in fact be a real physical effect -- high redshift systems might 
have more rotation or larger anisotropy \citep[see e.g.][]{vanderwel:ell.vsigma.evol}. 
Some process, e.g.\ dynamical heating from bars, or minor mergers, or clumpy 
star formation, could then vertically 
heat the systems by scattering stars and make them more round, while contributing little 
net mass or energy. Major axes would be little affected, while circular radii would 
increase. This effect could also occur owing to selection effects; 
if high-redshift samples (selected via a combination of Sersic indicex 
estimates and/or stellar population properties) include 
diskier systems or more early-type disks (e.g.\ Sa galaxies).
In either case, we include it as part of this category because 
the systems would appear to evolve along similar tracks, as both 
the apparent and real 
evolution in $R_{\rm circ}$ would involve no significant change 
in $\sigma$, $M_{\ast}$, or $\Sigma_{c}$. 
\\

{\bf Stellar Mass Uncertainties:} If the (uncertain) contribution of 
AGB stars to near-infrared light is large in young ellipticals 
\citep[ages $\lesssim2\,$Gyr, similar to the ellipticals at $z\sim2$;][]{kriek:drg.seds}, then 
the stellar mass $M_{\ast}$ as derived from commonly used stellar population 
models lacking a proper treatment of the TP-AGB phase \citep[e.g.][]{BC03} 
may be over-estimated by factors 
$\sim$a few \citep{maraston:ssps}. The difference will vanish as the populations 
age. This change in inferred $M_{\ast}$ will lead to apparent $R_{e}-M_{\ast}$ 
evolution ($M_{\ast}$ changes as fixed $R_{e}$); systems will appear less 
massive, but conserve $R_{e}$, $\sigma$, and profile shape. Likewise, 
redshift evolution in the stellar initial mass function, suggested (indirectly) by some 
observations \citep{hopkinsbeacom:sfh,
vandokkum:imf.evol,dave:imf.evol}, could yield a similar effect. 
\\

\begin{figure*}
    %\plotside{tracks_fixed_gal.ps}
    \plotside{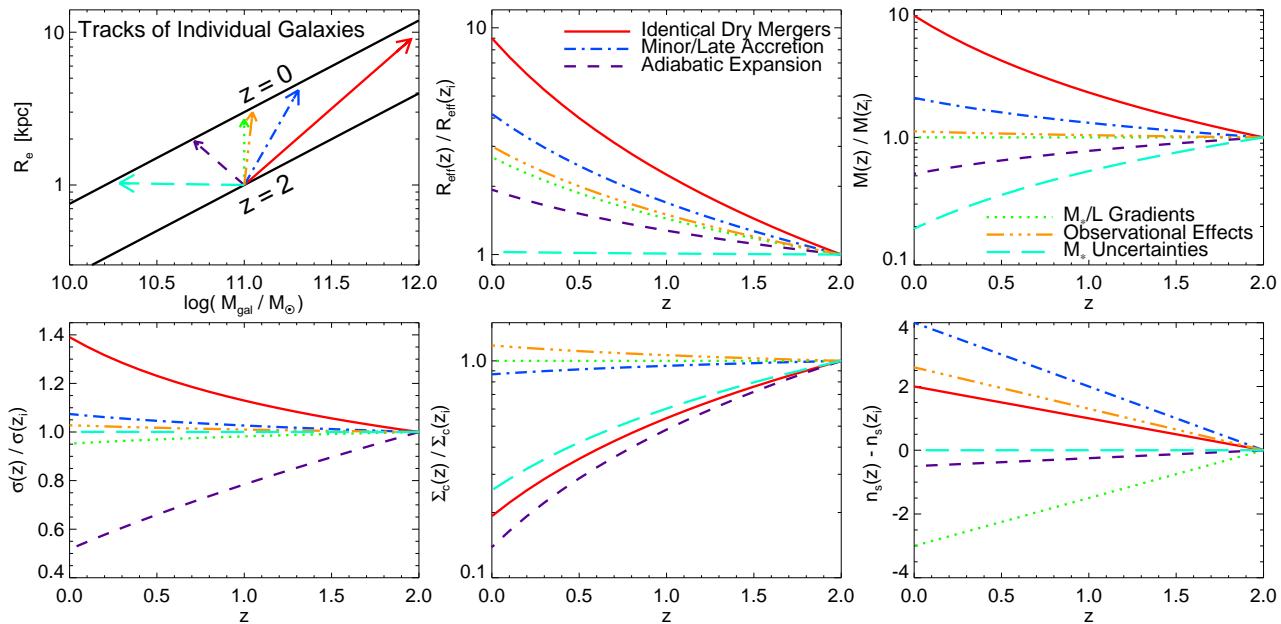}
    \caption{Evolution of a {\em fixed} population of galaxies, from fixed initial 
    conditions, according to different models. 
    {\em Top Left:} Tracks made by galaxies in the size-mass plane. 
    Solid line is the adopted ``initial'' 
    $z=2$ size-mass relation -- galaxies begin at $10^{11}\,\msun$ with 
    $R_{e}\sim1\,$kpc, and evolve until they lie on the $z=0$ size mass relation (dashed line). 
    We consider six models that could lead to apparent or physical size/mass evolution, 
    described in the text. In the following panels, we 
    see how this requires other changes in galaxy properties according to the model. 
    {\em Top Center:} Corresponding evolution in effective radius (relative 
    to the initial $z=2$ value) for each model,  
    for the fixed population of galaxies as a function of redshift. 
    The rate of evolution is such that they lie on the observed size-mass relation 
    at all times. 
    {\em Top Right:} Corresponding evolution in stellar mass, 
    required to produce the given size evolution. 
    {\em Bottom Left:} Corresponding evolution in the 
    velocity dispersion $\sigma$. 
    {\em Bottom Center:} Evolution in the 
    central/maximum surface stellar mass (not luminosity) density (i.e.\ some 
    $M_{\ast}\,kpc^{-2}$ averaged inside e.g.\ $\sim100\,$pc, 
    or containing $\sim1\%$ of the light; {\em not} the effective surface brightness). 
    {\em Bottom Right:} Evolution in apparent galaxy profile shape in e.g.\ 
    fixed rest-frame $B$-band, parameterized with the best-fit Sersic index $n_{s}$. 
    The different models clearly map out different average tracks in this space. 
    \label{fig:tracks.fixed.gal}}
\end{figure*}

Figure~\ref{fig:tracks.fixed.gal} plots how individual galaxies (or, since the 
history of an individual galaxy will be noisy, the median of a population of 
similar galaxies) evolve forward in 
time, according to these different models. We assume that at $z=2$, 
all systems ``begin'' on an observed $R_{e}-M_{\ast}$ relation similar to that 
inferred for observed systems at high mass -- specifically at 
an observed $10^{11}\,\msun$ with $R_{e}=1\,$kpc. They are evolved such that, at 
$z=0$, they will lie on the observed size-mass relation from 
\citet{shen:size.mass}. In each case, we assume that one and only one of the effects 
above operates, and we consider the strength of the effect to be arbitrary -- we make 
it as strong as necessary to evolve the systems onto the $z=0$ relation. Given 
this amount of evolution, though, we can quantify by how much e.g.\ the stellar 
mass must change. Likewise, we show 
how the velocity dispersions, the central/peak stellar 
mass density, and the best-fit Sersic index will change with time as the systems 
evolve towards the $z=0$ $R_{e}-\mstar$ relation along these tracks.

\begin{figure*}
    %\plotside{tracks_at_fixed_m.ps}
    \plotside{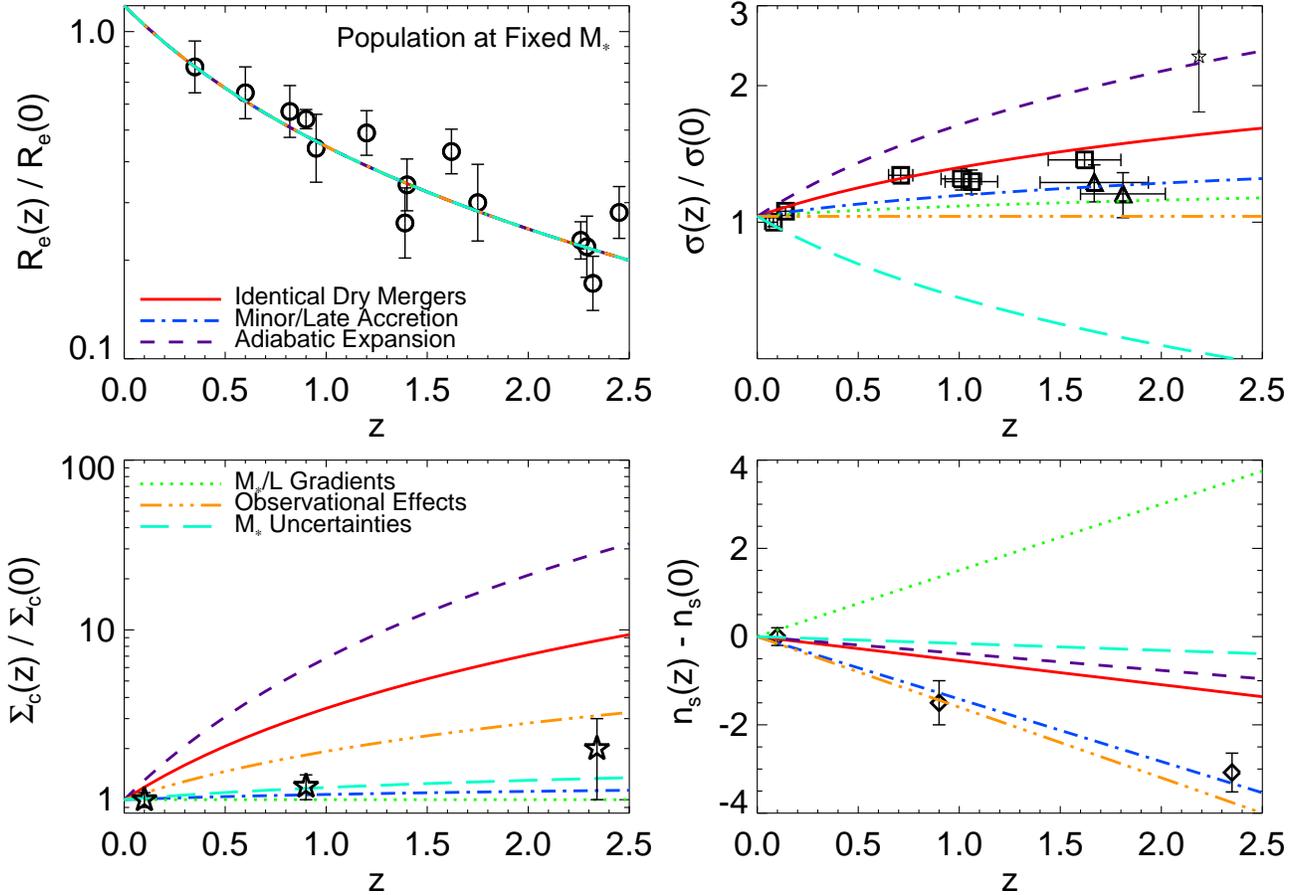}
    \caption{Evolution in properties of galaxies at fixed stellar mass, 
    given the different models from Figure~\ref{fig:tracks.fixed.gal}. 
    {\em Top Left:} Size evolution (median size observed 
    at redshift $z$, for galaxies of the same fixed observed stellar 
    mass, relative to the median size observed at $z=0$). We adjust each model to give 
    an identical size evolution at fixed mass, motivated 
    by the observational fits to the size evolution of 
    high-mass ($\gtrsim10^{11}\,\msun$) galaxies -- i.e.\ attribute the 
    observed size-mass relation evolution in each case {\em entirely} to just the one model. 
    We compare to the observed size evolution 
    \citep[points; from the compilation in][]{vanderwel:z1.compact.ell}. 
    In the following panels, we see the predicted consequences of 
    this for other quantities measured at fixed stellar mass. 
    {\em Top Right:} Velocity dispersion. Observations are compiled in 
    \citet{cenarro:sigma.of.highz.compact.ell} (squares; the $z>1.2$ point 
    comes from stacked spectra), as well as 
    \citet{cappellari:highz.gal.sigmas.and.mdyn} (triangles; also stacked spectra). 
    We also show the recent measurement from \citet[][star]{vandokkum:high.sigma.compact.gal}, 
    but note this is a single individual object. 
    {\em Bottom Left:} Central/peak stellar mass surface density ($M_{\ast}\,{\rm kpc^{-2}}$). 
    Observations are compiled in \citet{hopkins:density.galcores}. 
    {\em Bottom Right:} Profile shape/Sersic index. Observations 
    compiled from \citet[][$z=0$]{hopkins:cusps.ell}, \citet[][$z\sim1$]{vanderwel:z1.compact.ell}, 
    and \citet[][$z\sim2$]{vandokkum:z2.sizes}. 
    The different tracks in Figure~\ref{fig:tracks.fixed.gal} lead to different 
    predicted evolution in these quantities at fixed mass, if the observed 
    size evolution is attributed to each model in turn. 
    \label{fig:tracks.at.fixed.m}}
\end{figure*}

This is not directly comparable to what is done observationally, 
as the systems ``end up'' at different stellar masses.  
However, knowing how a galaxy population evolves forward in each 
model, it is straightforward 
to calculate how properties should evolve at fixed stellar mass, looking back on populations
with the same observed $\mstar$ at different redshifts. 
We show this in Figure~\ref{fig:tracks.at.fixed.m}. In detail, we force the 
$z=0$ galaxies to lie on all observed relations between structural properties 
and mass, and to match the observationally inferred evolution of 
$R_{e}(M_{\ast})$ for $\sim 10^{11}\,\msun$ galaxies, where the 
most dramatic size evolution has been seen.
\footnote{
Specifically, 
we adopt the observed $z=0$ relations: 
effective radius 
$R_{e} = 3.7\,{\rm kpc}\,(M_{\ast}/10^{11}\,\msun)^{0.56}$ \citep{shen:size.mass}, 
velocity dispersion 
$\sigma= 190\,{\rm km\,s^{-1}}\,(M_{\ast}/10^{11}\,\msun)^{0.28}$ \citep{hyde:scaling.relations.curvature}, 
central stellar mass surface density 
$\Sigma_{c} = 0.65\times10^{11}\,\msun\,{\rm kpc^{-2}}\,(M_{\ast}/10^{11}\,\msun)^{-0.15}$ 
\citep[fitted from the compilation in][]{hopkins:maximum.surface.densities}, 
and Sersic index (fitting the {\em entire} galaxy profile to a single Sersic index; 
as noted above this should be treated with some caution as the results depend on 
the fitted dynamic range)
%$\log{n_{s}} = 0.39 - 0.10\,(M_{B}+18) = 0.1755\,log(M_star/9.51236) +0.39 = 
% -> n_s=10^(0.39) * (M_star/9.51236)^(0.1755) = 4.478*(M_star/10^11)^(0.1755)
$n_{s} = 4.5\,(M_{\ast}/10^{11}\,\msun)^{0.18}$ \citep{ferrarese:profiles}. 
The central surface density $\Sigma_{c}$ must be defined: one can adopt 
either the extrapolation of the best-fit Sersic profile to $r=0$, or 
the average surface density within some small annulus (e.g.\ fixed physical 
$\sim 100\,$pc or fixed fractional $\sim R_{e}/50$); the details are discussed 
in \citet{hopkins:density.galcores} and \citet{hopkins:maximum.surface.densities}, 
but the differences are small for our purposes ($\Sigma$ is only a very 
weak function of $R$ in massive systems at these radii). 
For convenience we adopt the fixed $100\,$pc mean definition. 
We assume the specific form for the evolution of the size-mass relation 
$R_{e}(M_{\ast}\,|\,z) = R_{e}(M_{\ast}\,|\,z=0)\times (1+z)^{-1.5}$. 
This is convenient and yields a good fit to the observations shown in 
Figure~\ref{fig:tracks.at.fixed.m}, appropriate for massive galaxies, 
but the evolution may be weaker at lower masses (which has no effect 
on our conclusions). 
}
Integrating all populations backwards in time, we then reconstruct the properties at 
fixed stellar mass ($\sim10^{11}\,\msun$) according to each model. 

Clearly, these correlated properties can break 
degeneracies between different models tuned to reproduce the size distribution. 
For example, if adiabatic expansion were the explanation for the observed evolution, 
the velocity dispersions of 
$\sim 10^{11}\,\msun$ galaxies at $z\sim2-3$ would have to 
be $\sim2-3$ times larger than 
those today (because the system needed to have much more 
mass inside a small radius that was subsequently lost, and the expansion 
is homologous) -- 
i.e.\ typical $\sigma\sim 600\,{\rm km\,s^{-1}}$ at these masses. 
On the other hand, if 
minor/late accretion is responsible, the primary change has been the buildup of low-surface 
brightness ``wings'' in the profile, which contribute negligibly to $\sigma$ (leading to 
dispersions $\sim200-250\,{\rm km\,s^{-1}}$ at $z\sim2-3$). 

We compare with data compiled from recent literature. 
A number of measurements of the 
size distribution at fixed mass, hence the median size evolution, 
for massive ($M_{\ast}\sim10^{11}\,\msun$) galaxies are 
compiled in \citet{vanderwel:z1.compact.ell}; we adopt their compilation 
\citep[see][]{trujillo:compact.most.massive,longhetti:2007.z15.kormendy.relation,
zirm:drg.sizes,toft:z2.sizes.vs.sfr,cimatti:highz.compact.gal.smgs,vandokkum:z2.sizes,
franx:size.evol,rettura:2008.z1.sizes.clustervsfield,buitrago:highz.size.evol}. 
The Sersic indices of the high-redshift systems are presented 
in \citet{vandokkum:z2.sizes} and \citet{vanderwel:z1.compact.ell}; 
we compare these to the distribution of Sersic indices as a function of stellar 
mass at $z=0$ 
presented in \citet{hopkins:cusps.mergers,hopkins:cusps.ell,hopkins:cores}; 
itself a compilation from \citet{rj:profiles,jk:profiles,lauer:bimodal.profiles,ferrarese:profiles}.  
(Although \citealt{vanderwel:z1.compact.ell} note that a slightly higher Sersic index, 
similar to the $z=0$ observations, is also compatible with their sample). 
The high-redshift observations do not resolve the small radii needed to directly 
measure the central stellar mass density $\Sigma_{c}$; 
we adopt the estimates from \citet{hopkins:density.galcores} based on 
the best-fit profiles presented in \citet{vandokkum:z2.sizes} and 
\citet{vanderwel:z1.compact.ell}, extrapolating the fits from $\sim1-2\,$kpc 
inwards. In low-redshift systems this typically represents an upper limit 
(a factor $\sim1-2$ larger than the true central densities). 
The velocity dispersion measurements are compiled in 
\citet{cenarro:sigma.of.highz.compact.ell}; at the highest redshifts, they are 
measured therein from stacked spectra. Recently, 
\citet{cappellari:highz.gal.sigmas.and.mdyn} present a similar stacked 
analysis (and a re-analysis of the same objects) from $z=1.4-2$ and 
$z=1.6-2$, along with a couple of individual object measurements of 
$\sigma$; we show their results as well 
(from the stacked spectra; the individual detections are on the 
low-$\sigma$ end of the allowed range from the stack). We 
also show the recent detection presented in \citet{vandokkum:high.sigma.compact.gal}, 
of a very large $\sigma\sim500\,{\rm km\,s^{-1}}$ (albeit for a more massive 
$M_{\ast}=2\times10^{11}\,\msun$ system); however, we note that this is 
a single object (one of the brightest in the field), 
not a statistical sample, and the uncertainties in the measurement of 
$\sigma$ are large.

\section{Problems with Each Individual Model}
\label{sec:problems}

None of the models is ideal. 
Identical dry mergers can easily explain core creation in massive ellipticals 
(owing to the ``scouring'' action of binary supermassive BHs), but move systems 
relatively inefficiently with respect to the $R_{e}-\mstar$ relation. It requires a 
very large (order-of-magnitude) mass growth to get systems onto the $z=0$ 
size mass relation from $z=2$; this would yield too many $\sim 10^{12}\,\msun$ 
systems today \citep[relative to e.g.\ the local mass function from][]{bell:mfs}, 
given the number density of compact $z=2$ systems \citep{vandokkum:z2.sizes}. 
The predicted evolution in $\Sigma_{c}$ is much too strong, and that 
in $\sigma$ somewhat so. 

Minor/late mergers are a more efficient way to increase effective radii 
relative to the size-mass relation. This model fares best in 
Figure~\ref{fig:tracks.at.fixed.m} -- in fact, it is the only model that appears 
at least marginally consistent with all the observational constraints. 
However, the implied number of such mergers, to yield the full 
factor $\sim6$ evolution in $R_{e}(M_{\ast})$, is large -- 
larger than that predicted by cosmological models 
\citep{khochfar:size.evolution.model,hopkins:cusps.evol,
naab:size.evol.from.minor.mergers} or permitted by 
observational constraints implying $\sim1-2$\,major dry mergers 
(typical mass ratio $\sim$1:3) per system since $z=2$ 
\citep{vandokkum:dry.mergers,bell:dry.mergers,
lin:mergers.by.type,bridge:merger.fraction.new.prep,
darg:galzoo.merger.frac.by.morph,
darg:galzoo.merger.properties}. 
Figure~\ref{fig:tracks.fixed.gal} shows a typical $z\sim2$ compact 
system will have to increase its mass by a factor $\sim1.5-2$ 
to reside on the $z=0$ relation via this mechanism. This is 
not unreasonable, but it is still difficult, especially if in fact the high-redshift systems 
must ``overshoot'' the $z=0$ relation (such that older systems 
have larger $R_{e}$). 
This is ultimately reflected in the fact that models including 
just this effect tend to predict more moderate factor $\sim3-4$ size 
evolution \citep{khochfar:size.evolution.model,hopkins:cusps.evol}. 
It may be possible that the high-redshift systems reside preferentially 
in special environments with enhanced merger rates (relative to 
even other halos with the same mass and formation redshift); but 
we will show below that such an explanation is not necessary. 

Adiabatic expansion involves mass {\em loss}, so there is no 
issue with the mass function. However, the mass loss required 
to yield a large change in $R_{e}$ is correspondingly large, 
$>50\%$ of the $z=2$ mass. Even if all stars were just formed 
at this time, it is unlikely that stellar evolution could release this 
much mass (and it must be unbound, not simply recycled or heated). 
Given observed ages of the $z>2$ systems of 
$\sim0.5-1$\,Gyr \citep{kriek:drg.seds}, the expected subsequent mass 
loss is only $\sim 20\%$. One could posit that these systems have 
some other gas reservoir ``about to be'' blown out, but this appears to 
conflict with their low star formation rates and directly measured 
gas properties \citep[see e.g.][and references therein]{daddi05:drgs,
kriek:drg.seds,kriek:08.nir.spectroscopy.highz,
wuyts:irac.drg.colors,tacconi:smg.mgr.lifetime.to.quiescent}. 
Moreover, Figure~\ref{fig:tracks.at.fixed.m} 
clearly shows that the predicted evolution in $\sigma$ and 
$\Sigma_{c}$ is much larger than that observed. 

Mass-to-light ratio gradients should be present at $z=2$, if the 
observed stellar population gradients at low redshift are 
extrapolated back in time. However, using these to obtain more than 
a factor $\sim2$ in size evolution (the natural maximum in simulations) 
requires extremely low $M_{\ast}/L$ in the central regions (since 
there is effectively an upper limit to $M_{\ast}/L$ at large radii 
given by stellar populations with an age equal to the Hubble time). 
Such populations would require near-zero age, not the 
$\sim0.5-1\,$Gyr ages observed. It is also likely that dust, at such 
low age, would cancel out some of these effects, as in observed 
ULIRGs and recent merger remnants \citep[see e.g.][]{tacconi:ulirgs.sb.profiles,
rj:profiles}. Similar behavior (even frequent red cores in the merger 
and shortly post-merger phase) are also seen in 
gas-rich merger simulations \citep{wuyts:photometry.biases.mgrrem}.
Moreover, to the extent that this affects the fitted Sersic indices, 
accounting for the entire size evolution would 
imply higher $n_{s}$ at high-$z$ (since high-$n_{s}$ profiles 
have more concentrated central light), in conflict with the observations 
in Figure~\ref{fig:tracks.at.fixed.m}. 

Attributing the entire evolution to incorrect stellar mass estimates 
appears similarly unlikely. It requires 
invoking something more than just the known differences between e.g.\ the 
\citet{maraston:ssp.effects} and \citet{BC03} stellar population models -- applied to the observed 
$z=2$ systems with the best present data, these give only factor $\sim1.4$ difference 
in stellar mass \citep{wuyts:irac.drg.colors}. 
In any case, the required evolution in $\sigma$ clearly disagrees with the observations. 

Observational biases from e.g.\ profile fitting appear marginally consistent 
with the constraints in Figure~\ref{fig:tracks.at.fixed.m}. However, attempts 
to calibrate such effects typically find they lead to bias in high-redshift 
sizes at the factor $\sim2$ level \citep{boylankolchin:mergers.fp,
hopkins:density.galcores} or smaller \citep{vanderwel:z1.compact.ell}; 
not the factor $\sim6-10$ desired. 
Stacking the high-redshift data also appears to yield similar sizes, so it is 
unlikely that a very large fraction of the galaxy mass lies at radii not 
sampled by the observations \citep{vanderwel:z1.compact.ell}. 
In fact, experiments with hydrodynamic simulations suggest that, if 
anything, biases in fitting Sersic profiles may lead to {\em over}-estimates 
of the high-redshift sizes (S.\ Wuyts et al., in preparation), and the 
higher dissipational fractions involved in forming compact ellipticals can yield 
sharp two-component features that bias the fits to higher Sersic indices and 
corresponding effective radii. 
Moreover, it is difficult to invoke these effects to explain the 
observed $R_{e}(M_{\ast})$ 
evolution at lower stellar masses, where $z=0$ Sersic indices are 
relatively low so there is less of an effect from 
the extended tails of the light distribution. 

It is also straightforward to check whether or not evolution in 
the shapes of ellipticals, at otherwise fixed major axis radii and structural 
properties, accounts for the observed evolution (via use of the 
circularized radius). Several of the studies discussed here present 
not just the masses and effective radii of their systems, but also their 
major and minor axis lengths; we compile these 
from \citet{trujillo:size.evolution,trujillo:compact.most.massive,
lauer:bimodal.profiles,vandokkum:z2.sizes,damjanov:red.nuggets,jk:profiles}, 
a sample from $z=0-2.3$, and use this to compare the evolution in major 
and minor axis lengths. Restricting our analysis to major axis lengths alone, 
we find that the evolution is slightly weaker than that using circularized 
radii (in other words, there is some increase in the median ellipticity of the 
samples with redshift); however, the effect appears to be relatively small, 
accounting for a factor $\sim(1+z)^{-(0.2-0.5)}$ in evolution (i.e.\ $\sim20\%$ of the 
total size evolution). This is comparable to the effects anticipated from 
simulations (see \S~\ref{sec:fgas.evol}). 

As observations 
improve, it appears unlikely that these effects 
can account for the full evolution, although 
it may be important (altogether) 
at the factor $\sim1.5-2$ level and cannot be entirely  
ruled out by the constraints in Figure~\ref{fig:tracks.at.fixed.m} alone.

\section{An A Priori Cosmological Model}
\label{sec:mixed}

\begin{figure*}
    %\plotsidesmall{demo_sim.ps}
    \plotsidesmall{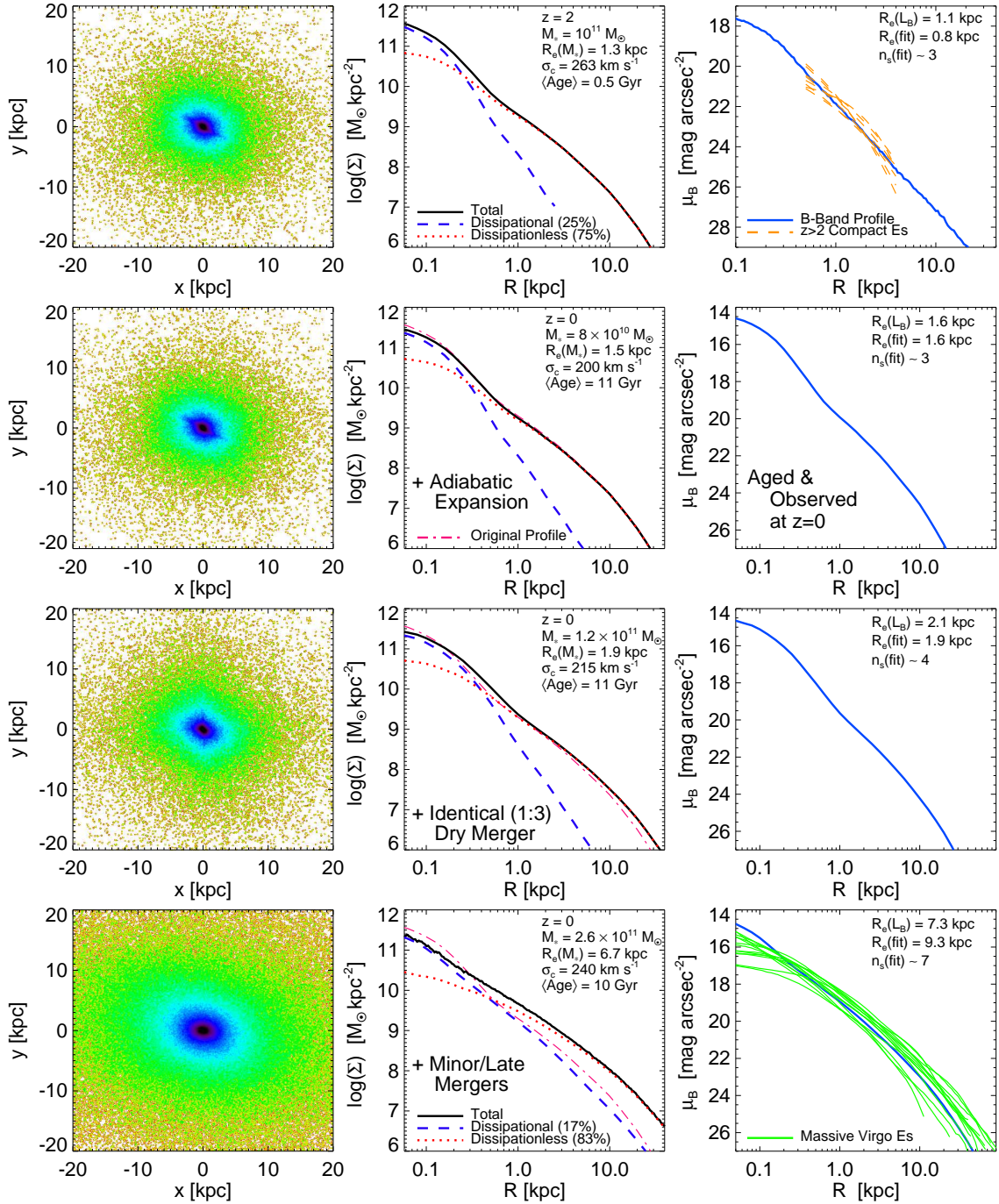}
    \caption{Illustration of a typical history from cosmological simulations, 
    realized in high-resolution hydrodynamic simulations, for a massive, 
    early-forming spheroid. 
    {\em Top:} Image of stellar surface density ({\em left}), 
    and axially-averaged projected stellar surface mass density profile 
    ({\em center}; solid) after the spheroid first forms in a $z\sim3$ gas-rich 
    ($f_{\rm gas}\sim 30-40\%$) merger. Physical properties 
    as seen at $z=2$ are shown. We also show the profile decomposed into 
    the dissipational (post-starburst) stars (with mass fraction $\sim25\%$) 
    and dissipationless (pre-merger disk) stars. We construct the rest-frame 
    $B$-band profile ({\em right}) with half-light radius $R_{e}(L_{B})$, 
    and a mock $z=2$ observation (with mock resolution, PSF, and surface brightness 
    limits) with subsequent best-fit Sersic function effective 
    radius $R_{e}(\rm fit)$ and index $n_{s}$. We compare to the observed $z>2$ 
    compact spheroid profiles from \citet{vandokkum:z2.sizes}, over the 
    dynamic range outside the PSF and above the background limits. 
    {\em Second from Top:} Same, after adiabatic expansion and stellar evolution 
    are allowed to operate. Stars lose appropriate mass for each stars age/metallicity 
    evolved forward to $z=0$, which then virializes; the resulting mass profile is compared 
    to the original ($z=2$) profile ({\em center}). We also re-construct the $B$-band profile, 
    with the stellar populations aged in this way (allowing for $M_{\ast}/L$ evolution 
    per the observed/simulated properties), and re-fit it assuming image depth 
    comparable to local observations. 
    {\em Second from Bottom:} Same, including a single (mass ratio 1:3) ``identical'' 
    dry merger (high-redshift dry merger with similarly gas-rich disk or compact elliptical) 
    in the history. 
    {\em Bottom:} Further adding a small series of later major and minor 
    mergers of less-dense systems (typical lower-redshift material in disks and 
    spheroids). Since this material has lower dissipational content (forming 
    from lower-redshift, more gas-poor disks), the net dissipational fraction goes down, 
    and extended envelopes preferentially build up. We compare the $B$-band profile 
    to a few observed massive spheroids in Virgo \citep{jk:profiles}. 
    \label{fig:demo.sim}}
\end{figure*}

In principle, {\em all} of these effects can occur. We therefore 
consider a cosmological model for galaxy growth in which they
are all included.  We follow a ``typical'' 
massive spheroid formed at $z\sim3$ from the cosmological model 
in \citet{hopkins:cusps.evol}, using the high-resolution hydrodynamic 
simulations presented in \citet{cox:kinematics,robertson:msigma.evolution,
hopkins:cusps.mergers}. Those simulations include self-consistent 
models for star formation and black hole growth, and feedback 
from both, that enable the stable evolution of even very gas rich systems
\citep{springel:models,springel:spiral.in.merger}. 
They survey a wide range of parameter space, in both gas-richness, 
progenitor redshift, and structural properties of the merging galaxies, 
that make them well-suited for the experiment below. 
We illustrate the evolution of the system in Figure~\ref{fig:demo.sim}.

\begin{figure*}
    %\plotside{tracks_at_fixed_m_mixed.ps}
    \plotside{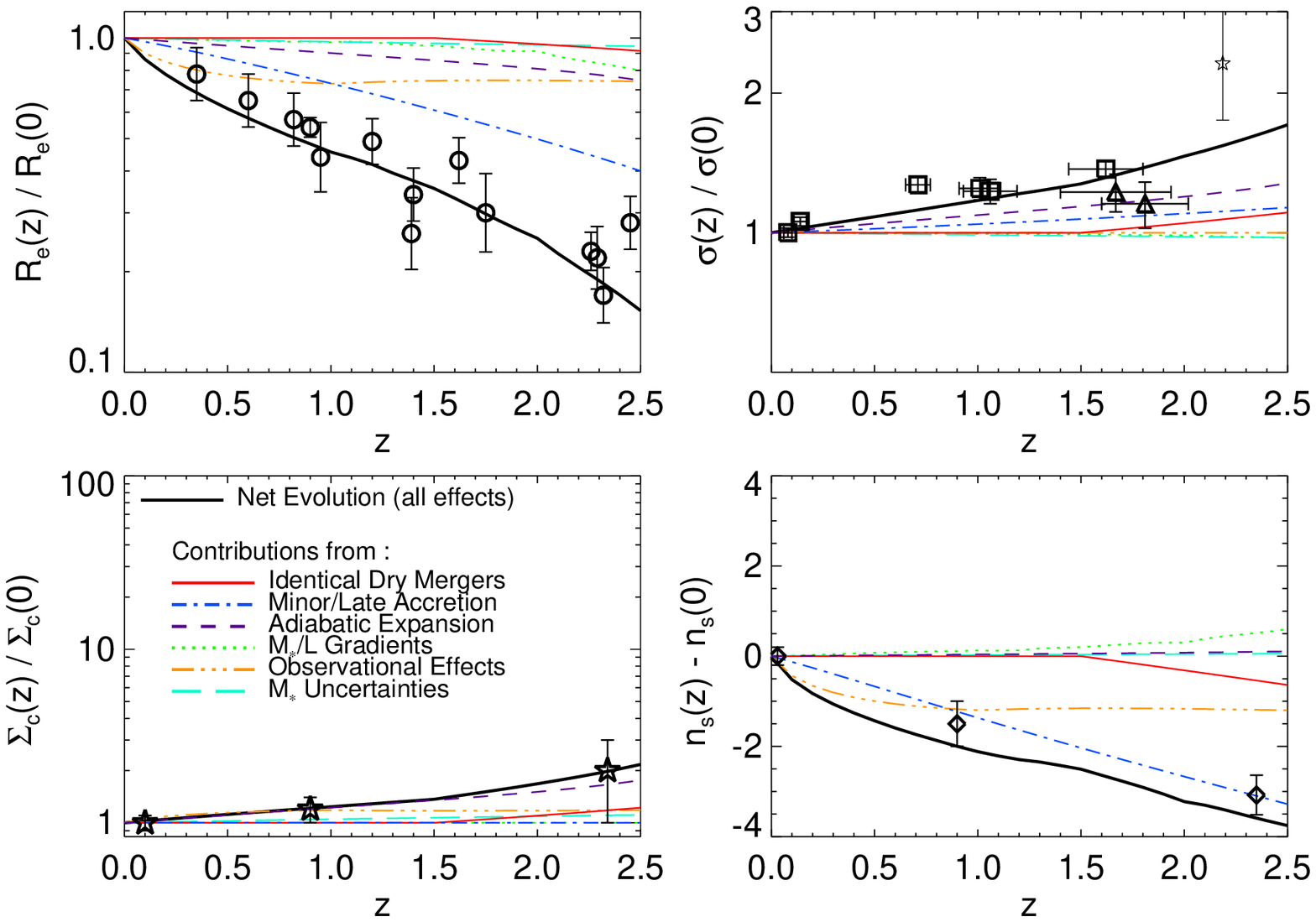}
    \caption{Predictions of a cosmological model where all 
    the effects in the text contribute to size evolution (as in Figure~\ref{fig:demo.sim}). 
    In each, we consider the prediction for the net evolution (black solid line) 
    as in Figure~\ref{fig:tracks.at.fixed.m}. We show the contribution to evolution 
    in each track from each of the individual effects described in the text. 
    The ``late/minor merger'' channel dominates growth, yielding a factor $\sim3-4$ size 
    evolution with small (factor $\sim2$) mass growth, but each of the other effects 
    here contributes a small $\sim20\%$ additional effect. Together, this gives an 
    additional factor $\sim2-3$ needed to explain the observed trends with redshift 
    given cosmologically realistic merger/growth rates and without violating 
    constraints from stellar mass functions or other observations. 
    \label{fig:mixed.model}}
\end{figure*}

The system first becomes a spheroid in a $z=3$ merger of two 
$\sim L_{\ast}$ disks, with typical gas fractions for their mass and redshift. 
We specifically choose as representative one of the simulations 
described in detail in \citet{hopkins:cusps.fp} (typical of mergers with such 
gas fractions, with common orbital parameters and halo properties). 
The simulation ``begins'' at $z=3$ (and completes 
at about $z\sim2.4$), and models an equal-mass 
merger of two identical massive disks, each with a gas fraction (at the time 
of the merger) of $30-40\%$, representative of observations near these 
masses \citep{erb:lbg.gasmasses}. The remnant formed 
in such a gas-rich merger is very compact (since this $\sim30-40\%$ is 
entirely channeled into the central starburst, the effective radius -- $50\%$ 
mass radius -- lies just outside the compact central starburst region). 
Since the halo mass has crossed the ``quenching threshold'' where 
cooling becomes inefficient 
\citep[$\sim 10^{12}\,\msun$; see][]{keres:hot.halos} with this merger, 
and the observations indicate that these high-redshift systems 
do not form many stars from their formation redshifts to today (likewise 
massive ellipticals today have not formed significant stars since $z\sim2-3$), 
we turn off future cooling, and evolve the system passively for a short time 
to $z=2$ (to allow it to relax and redden \citep{springel:red.galaxies}). 
Figure~\ref{fig:demo.sim} ({\em top}) shows the resulting stellar 
mass surface density profile, together with some salient parameters. 
The system has 
a true stellar mass of $M_{\ast}=1.0\times10^{11}\,\msun$, a mass-weighted central 
velocity dispersion of $\sigma=260\,{\rm km\,s^{-1}}$, a central/peak stellar surface 
mass density of $\Sigma_{c}=2\times10^{11}\,\msun\,{\rm kpc^{-2}}$, 
and a stellar effective radius $R_{e}=1.3\,$kpc. 
Note that $\sigma$, $\Sigma_{c}$, and $R_{e}$ are projected quantities; 
here, we sample the system at each time by projecting it along $\sim100$ 
lines of sight, uniformly sampling the unit sphere, and quote the 
median, but the sightline-to-sightline variance in these quantities 
is small \citep[$\lesssim0.1\,$dex; see][]{hopkins:cusps.mergers}. 

We then construct the light profile of the system at $z=2$. 
The stellar population properties (ages and metallicities) are determined 
self-consistently from the star formation and enrichment model in the 
simulation; dust properties are computed self-consistently from the 
simulation gas and sub-resolution ISM model, following the methodology 
in \citet{hopkins:lifetimes.letter,hopkins:lifetimes.methods} (but at this time, 
the gas is depleted or in the hot halo, so this is a relatively small correction). 
The $B$-band light weighted stellar population age at this 
time is $\sim500\,$Myr, very similar to that inferred from the 
observed high-redshift systems \citep{kriek:drg.seds}. 
Figure~\ref{fig:demo.sim} shows the true rest-frame 
$B$-band light profile, and corresponding parameters. 
There is a gradient in the stellar mass-to-light ratio, owing to the 
recently-forming starburst populations; as a result, the 
profile shape and effective radius are slightly different in 
$B$-band than in stellar mass. The $B$-band effective radius is 
$1.1\,$kpc; $M_{\ast}/L$ gradients in this system are not negligible, 
but contribute only a $\sim20-30\%$ effect in the size. 

We compare this directly to the observed best-fit (de-convolved) profiles 
of $z>2$ galaxies from \citet{vandokkum:z2.sizes}, shown over the range 
where the observations are well-sampled 
and not severely affected by the PSF (roughly $\sim1-4\,$kpc). 
The two agree reasonably well. 

But in order to estimate the observational effects of e.g.\ limited dynamic 
range and fitting, to compare directly to the observed best-fit sizes, 
we construct a (toy model) mock observation of the system. 
We consider an image in the rest-frame $B$-band, with 
instrument quality comparable to HST imaging, 
and a simple representative PSF,\footnote{For the sake of 
generality, we adopt a simple Gaussian PSF with 
$1\,\sigma$ width $=0.5\,$kpc (FWHM$=1.2\,$kpc), 
representative of the best-case resolution of 
most HST observations at $z\gtrsim0.5$.} and (simple noise) 
sky background, and fit it in a 
manner designed to mimic the observations (here a one-dimensional 
fit to the circularized profile allowing for varying ellipticity and 
isophotal twists with radius).  
The mock ``observed'' best-fit Sersic parameters 
are shown in Figure~\ref{fig:demo.sim}. 
The best-fit observed $R_{e}$ is $0.8$\,kpc, with a fitted Sersic index 
of $n_{s}=3$ ($=0.9$\,kpc if we fix $n_{s}=4$). 
Observational biases from missing some of the outermost light and, 
more importantly, having the best sampling of the profile 
in the central regions where the profile shape is {\em not} 
the same as in the outer regions (thus leading to a fit that reflects 
that central portion), lead to a slightly smaller fitted size than the 
true $B$-band size. But the effect is small, only $\sim10-20\%$. 
Simple experiments suggest that the presence of such an 
effect depends on the exact image construction and fitting methodology, 
and a more thorough comparison of simulated profiles and high-redshift 
data will be the subject of future work
(S.\ Wuyts, in preparation); the important point here is that the effect being 
small at early times means that it is not especially important whether or not 
such a bias exists, in explaining the overall size evolution of the system. 

Note that the system here is relatively round, with a median ellipticity 
(over an ensemble of random sightlines) of $\epsilon=1-b/a\approx0.2-0.3$. 
Thus, if we were to adopt the circularized radius instead of the 
circular half-light radius or major axis radius, we would obtain a smaller 
radius by a factor of $\sqrt{b/a}\approx0.84-0.89$. Although a small correction, 
this is comparable to the other effects here and is not negligible -- even 
in a relatively round system viewed from a random viewing angle, 
the circularized radius yields another $\sim10-20\%$ smaller apparent size. 
In more flattened systems, or from the more extreme viewing angles for this 
particular simulation, the effect can be as large as $\sim30-40\%$.

We can also consider the role of stellar mass uncertainties: 
the maximum of this effect is given by the following. We model the 
``true'' stellar population parameters with the \citet{maraston:ssps} 
stellar population models (including a large contribution to the NIR 
light from AGB) stars; we then fit the mock photometry with the 
\citet{BC03} models. If we do so, the best-fit stellar mass 
obtained from the mock image is 
$M_{\ast}=1.2\times10^{11}\,\msun$. In other words, the 
maximum of this effect is a $\sim20\%$ bias in stellar mass, 
which given the observed correlation $R_{e}\propto M_{\ast}^{0.6}$, 
would lead to just a $5\%$ effect in the apparent size-mass 
relation. 

As the system evolves in time, the stellar mass-to-light ratio gradients 
become progressively weaker 
\citep[the system becomes uniformly old, and moreover metallicity 
gradients increasingly offset age gradients; see][]{hopkins:cusps.ell}, 
as does the difference between the stellar mass inferred from different stellar 
population models, 
and observed at $z=0$ with the quality of the best observations, biases 
in the fitted size will vanish as well. 
In fact, because of metallicity gradients (higher metallicity towards the 
self-enriched starburst material at the center) yielding a slightly 
red core \citep[typical in $z=0$ massive ellipticals; see][]{peletier:profiles,caon90,
kuntschner:line.strength.maps,sanchezblazquez:ssp.gradients}, 
the $B$-band $R_{e}$ here is about $\sim10\%$ larger than the 
stellar mass $R_{e}$ at late times. 
If this is all that would happen, of course, the observed $B$-band size 
will simply converge or slightly 
over-estimate the true stellar-mass size; giving only a 
factor $1.5-2$ evolution and yielding a system that is still much 
smaller than comparable ellipticals today. 

But, as the system ages, other effects will contribute. First, 
consider adiabatic expansion. Modeling the stellar populations 
as with the \citet{BC03} models, we know how much stellar 
mass loss occurs as a function of time. Since observations indicate the 
systems have evolved passively since high redshift, we assume that this 
mass lost is heated \citep[easily accomplished by e.g.\ shocks or AGN 
feedback, see e.g.][]{ciottiostriker:recycling} and contributes to the 
hot, virialized halo in hydrostatic equilibrium \citep{cox:xray.gas}.
Since all that matters for 
the size of the system is that most of the gas not be in the compact 
central regions, it makes no difference whether we assume this 
model or that the gas is expelled completely. Likewise, it makes no 
difference if we assume all the remaining gas at the end of the 
merger is rapidly expelled (most of the gas has, at that point, been depleted 
by star formation; the much larger gas reservoir for adiabatic expansion 
comes from stellar mass loss). At each time, we then adiabatically 
expand the system according to the mass lost following 
\citet{gnedin:adiabatic.contraction}. 
Although, in principle, up to $\sim50\%$ of the stellar mass is 
recycled by stellar mass loss, much of this occurs when the stellar 
populations are very young (to rough approximation, mass loss 
rates at late times decline as $\sim t^{-1}$). Given the \citet{BC03} 
models and the distribution of stellar population ages already in place 
at $z=2$, the stars here will lose only $\sim20\%$ of their mass 
over the remainder of a Hubble time. At $z=0$ then, the net effect 
of adiabatic expansion (alone) would be to leave the profile shape the same, 
reduce the stellar mass to $0.8\times10^{11}\,\msun$, 
inflate the effective radius to $1.5\,$kpc, 
and decrease the velocity dispersion and 
to $\sigma=200\,{\rm km\,s^{-1}}$ 
(note that $\sigma$ has contributions from dark matter and 
large radii, so is not quite as strongly affected as might be expected). 

The system experiences a significant merger history, involving 
several different mergers. Here, we select a representative merger 
history from the cosmological models in \citet{hopkins:merger.rates}\footnote{
The merger rates from this model can be obtained as a function of galaxy 
mass, redshift, and merger mass ratio from the ``merger rate calculator'' 
script publicly available at 
\mergercalcurl.
}, 
where these are discussed in detail, for a galaxy that is already a 
massive $\sim10^{11}\,\msun$ at $z>2$. 
We simulate that merger history at high-resolution, and show the results in 
Figure~\ref{fig:demo.sim}. 

For simplicity, and to correspond to the model classification scheme used throughout 
this paper, we divide the mergers in the cosmological model into ``identical dry mergers'' 
and ``minor/late mergers'', and consider each in turn (note that, for the final remnant, the 
exact time-ordering of the mergers does not make a significant difference). 
First, consider the ``identical dry merger'': around $z\sim2$, the system experiences 
a (marginally) major merger (mass ratio $\approx$1:3) with another spheroid 
that was, itself, also formed in a $\sim30-40\%$ gas merger 
(again, since this is still at high redshift, these numbers are typical; more gas-poor 
systems are not typical), and should therefore be similarly dense. 
We model this by considering a 1:3 merger with a similar spheroid formed, itself, 
in an equally gas-rich merger. 
The effective radius expands $R_{e}\propto M_{\ast}$, as expected, 
and the velocity dispersion increases very slightly (it is not perfectly constant 
because there is some preferential transfer of energy to the least-bound outer material 
instead of the central regions). 
The net effect is a relatively small increase in size ($R_{e}=1.9\,$kpc) and 
little growth relative to the size-mass relation. 

However, at the same time and, increasingly, at later times, the system 
experiences a number of minor and major ``late'' (less dissipational or 
less dense) mergers. 
Specifically, in this system, we consider just those above a mass 
ratio 1:10 (more minor mergers contribute a negligible $\sim5\%$ to the 
final mass). In general, \citet{hopkins:merger.rates} show that $>L_{\ast}$ 
systems have enhanced merger rates, but the mergers become progressively 
more minor (on average) as the system grows in mass. 
This simply reflects the fact that most of the mass in the Universe is in 
$\sim L_{\ast}$ systems, so the growth will be dominated by
mergers of those systems (which are minor when the system is 
$\gg L_{\ast}$). 
In the particular Monte Carlo history modeled here, this is manifest in the 
series of ``minor/late mergers.'' 

The system, around $z\sim0.5-1$, experiences its 
last major merger, with a more gas-poor disk of mass $\sim0.5\, L_{\ast}$ 
(a roughly 1:3 merger). The disk has a gas fraction of $\sim 20\%$ at the time 
of merger -- still not negligible, but significantly lower than the $\sim 40\%$ 
that made the original spheroid. Therefore it contributes proportionally more 
dissipationless (stellar disk) material, which is low-density relative to the 
dissipational (gas/starburst) material, and therefore preferentially builds up the 
``wings'' of the profile. 
Similarly, this is accompanied and followed by a sequence of a $\sim$1:5 merger, a $\sim$1:6 
merger, and a $\sim$1:10 merger, spanning redshifts $z\sim0-1.5$. 
Whether those secondary galaxies are modeled as disks or spheroids makes little 
difference -- the key is that they have the appropriate gas or dissipational fractions for 
their mass and redshift ($\sim 10-20\%$). 
In fact, we also find that the precise time ordering or even the mass ratios of the 
mergers is not a significant source of uncertainty in the predicted mass profile. 
To lowest order, the important quantities are simply the total 
mass added to the ``dissipationless'' low-density envelope (i.e.\ material contributed 
by lower-redshift, low-density stellar disks, whether or not these come in 
minor units or are pre-processed into spheroids) versus the mass 
added to the ``dissipational'' starburst component 
\citep[for a detailed discussion, see][]{hopkins:cores,hopkins:cusps.evol}. 
We appropriately account for adiabatic mass loss in each galaxy as 
described above, and construct a mock image of the 
final remnant as observed at $z=0$. 

Allowing for this merger and accretion history, the system has grown 
by a moderate (albeit non-trivial) factor $\sim2.5$ in stellar mass. 
This is small enough that there is no conflict with observed stellar 
mass function constraints: even if every such compact $z>2$ 
massive ($M_{\ast}\gtrsim10^{11}\,\msun$) passive galaxy grew by such a 
factor, it would account for only $\sim20-50\%$ of the $z=0$ 
$M_{\ast}\gtrsim2-3\times10^{11}\,\msun$ spheroid population 
\citep[see e.g.][]{perezgonzalez:mf.compilation,marchesini:highz.stellar.mfs,
vanderwel:size.numden.massive.ell}. 
The velocity dispersion grows by a small amount for the same 
reasons as in the ``identical dry merger'', but is still 
moderate for the total stellar mass, $\sigma\approx240\,{\rm km\,s^{-1}}$. 
But the effective radius has grown by a factor $\sim6-10$. 
The stellar mass and $B$-band effective radii are now $\sim 7\,$kpc, 
and the fitted $R_{e}$ is slightly larger, owing to the large best-fit Sersic 
index $n_{s}\sim7$ (which itself owes to the extended envelope of 
low-density material acquired via these late/minor mergers). 
The system is also extremely round at this point -- there is only a $\sim5\%$ 
difference between the circular half-light radius and the ``circularized'' radius. 

We compare the predicted final profile to the observed light profiles of the 
$\sim10$ most massive Virgo ellipticals 
\citep[from][]{jk:profiles}, spanning a mass range 
$\sim 10^{10.6-11.7}\,\msun$. The profile appears quite typical. 
Note that although the central density inside $\sim 100\,$pc may be 
slightly high, by a factor $\sim1.5-2$, this is still well 
inside the scatter in central profile shapes
typical in larger samples of ellipticals \citep[see e.g.][]{lauer:bimodal.profiles}. 
Moreover, our simulations do not include any process of core ``scouring,'' 
by which BH-BH mergers are expected to eject stars inside precisely these 
radii ($\sim10-100\,$pc, for galaxies with the masses of our final remnant) 
and convert initially ``cuspy'' nuclear profiles (formed in gas-rich starbursts 
as those here) into ``cored'' profiles. Including a toy model for 
such a process following \citet{milosavljevic:core.mass}, 
we find that the predicted profile is fairly typical. 

Therefore, allowing for the combination of all the effects considered here, 
we find it is possible to account for even order-of-magnitude size evolution 
in the most massive, early-forming compact spheroids. 
The most important contribution to that evolution is the 
``minor/late merger'' channel, as expected based on our comparisons 
in \S~\ref{sec:models}. But as noted in \S~\ref{sec:problems}, given 
cosmologically realistic merger histories and observational 
constraints on galaxy merger/growth rates, this channel is only expected 
to contribute a factor $\sim3-4$ in size growth. 
The remaining factor $\sim2-3$, we find, comes from a combination of 
the other effects considered here: $M_{\ast}/L$ gradients, 
observational effects, adiabatic expansion, and identical dry mergers. 
Each of these effects contributes only a relatively small 
$\sim10-30\%$ effect to the size evolution -- but together, this yields the 
net factor $\sim2-3$ additional evolution needed to reconcile the 
observations at $z>2$ and $z=0$. 

Given the combination of these effects, we can now attempt to predict 
the evolution of properties at fixed mass, as in Figure~\ref{fig:tracks.at.fixed.m}, 
but for realistic systems where all of these effects act at different levels. 
We use the Monte Carlo merger populations from \citet{hopkins:merger.rates}, 
together with the simple analytic fitting functions for how sizes 
and velocity dispersions change in mergers of a given 
mass ratio, size ratio, and gas fraction, from \citet{hopkins:cores,
hopkins:cusps.fp,hopkins:cusps.evol}, and we model the 
effects of adiabatic expansion in each system as described above. 
The magnitude of observational bias as a function of redshift we estimate 
following the (very simplified) methodology in \citet{hopkins:density.galcores}, 
and mass-to-light ratio gradients are modeled as a function of time 
by using the mean scalings of stellar population gradient strength with 
post-merger gas fraction and age from simulations in \citet{hopkins:cusps.ell}. 
The details are not particularly important; we could assume all these 
effects scale with time and redshift exactly as in the specific experiment 
shown in Figure~\ref{fig:demo.sim} and we would obtain a similar answer. 

Figure~\ref{fig:mixed.model} shows the results. We plot how the median size, 
velocity dispersion, central stellar mass surface density, and best-fit 
Sersic index scale with redshift at fixed mass, given this combination of effects. 
In each case they agree well with the observations, and are similar to the 
``minor/late merger'' track in Figure~\ref{fig:tracks.at.fixed.m}, as expected. 
We also show the contribution to each effect from each of the different 
channels modeled. As before, although late/minor mergers dominate, 
the other effects together contribute comparably, and no single one of those 
effects is especially more important than the others.

\section{Implications for the Evolution in BH-Host Galaxy Correlations}
\label{sec:bh}

\begin{figure}
    %\plotonesmall{BH_tracks.ps}
    \plotonesmall{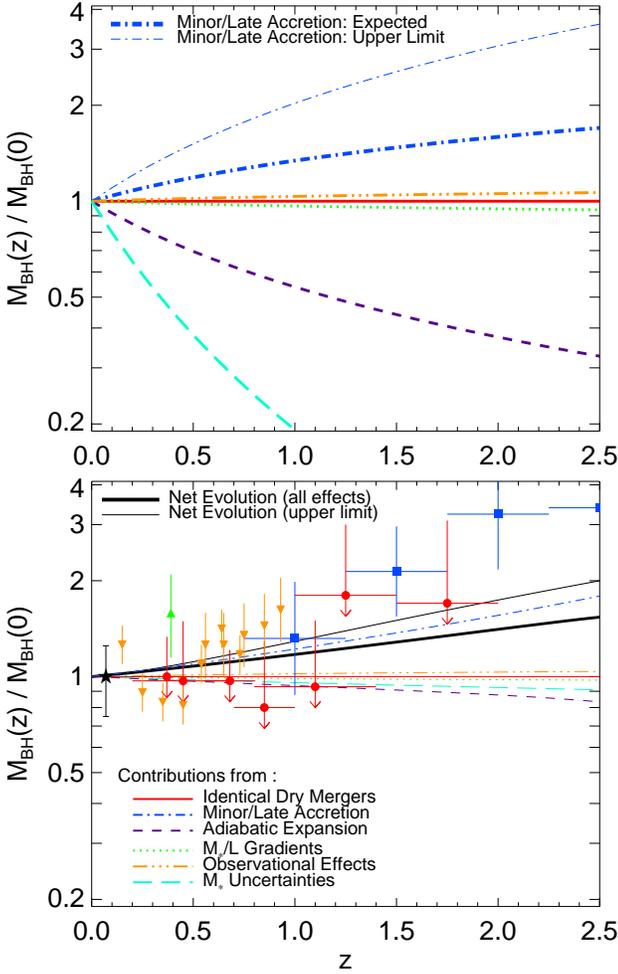}
    \caption{Predictions for the evolution in BH mass at fixed host galaxy stellar 
    mass, in the style of Figure~\ref{fig:tracks.at.fixed.m} ({\em top}) and 
    Figure~\ref{fig:mixed.model} ({\em bottom}), for the same models. 
    %{\em Top:} $M_{\ast}/L$ gradients and observational/seeing effects involve 
    %passive evolution of observationally inferred radii, hence no BH-host evolution; identical dry mergers 
    %move systems strictly along the $M_{\rm BH}\propto M_{\rm bulge}$ 
    %relation. Adiabatic expansion and stellar mass uncertainties 
    %require the bulge being (or appearing) more massive at high redshifts, 
    %hence predict lower $M_{\rm BH}/M_{\rm host}$ at high-$z$. 
    Observational effects, $M_{\ast}/L$ gradients, and identical dry mergers 
    all give no evolution; adiabatic expansion and $M_{\ast}$ uncertainties 
    lead to higher (real or inferred) bulge masses at high-$z$, hence lower $M_{\rm BH}/M_{\rm bulge}$. 
    Accounting for subsequent BH accretion only makes the predicted evolution more negative. 
    Minor/late mergers allow for evolution towards larger $M_{\rm BH}/M_{\rm bulge}$ 
    at high-redshift; the bulge ``catches up'' via later mergers with e.g.\ gas-poor disks. We show 
    both the upper limit of such evolution (assuming that the late/minor mergers 
    contribute {\em no} BH mass) and the cosmologically expected evolution 
    (given BH masses and merger histories from hydrodynamic simulations). 
    {\em Bottom:} Same, from the full model of \S~\ref{sec:mixed}, with 
    contributions from each effect. 
    Observational constraints are compared, from 
    \citet[][black star]{haringrix}, 
    \citet[][blue squares]{peng:magorrian.evolution}, 
    \citet[][green triangle]{woo06:lowz.msigma.evolution,treu:msigma.evol}, and 
    \citet[][orange inverted triangles]{salviander:midz.msigma.evol}. 
    Upper limits derived from 
    observations of the spheroid mass density and the fact that BH 
    mass cannot {\em decrease} with time are also 
    shown, from \citet[][red circles]{hopkins:msigma.limit}. 
    The model agrees reasonably well, but the observations are systematically uncertain, 
    and various biases probably affect the determinations at high redshifts 
    (see text). 
    %In any case, {\em only} the minor/late mergers channel allows for 
    %positive evolution in $M_{\rm BH}/M_{\rm host}$. 
    \label{fig:bhs}}
\end{figure}

Given these models, we briefly consider their implications for the evolution in the 
correlations between black hole (BH) mass and various host galaxy properties. 

In the models which do {\em not} involve mergers, considered here, the 
expected evolution of BH-host correlations is particularly simple. We perform 
the same exercise as in \S~\ref{sec:models}, and require that 
all the models reproduce the observed $z=0$ BH-host relations. 
Specifically, we enforce the $z=0$ relation between BH mass and bulge 
stellar mass \citep[$M_{\rm BH}=0.0014\,M_{\ast}$;][]{haringrix}, although
adopt instead the $z=0$ $M_{\rm BH}-\sigma$ relation 
\citep[$M_{\rm BH}=10^{8.13}\,(\sigma/200\,{\rm km\,s^{-1}})^{4.02}$;][]{tremaine:msigma} 
makes no difference. Since the BH cannot {\em decrease} in mass with time, 
this sets at least a limit on the evolution. We evolve the systems backwards in 
time, and predict the observed BH mass at fixed galaxy properties. 

In the case of stellar mass-to-light ($M_{\ast}/L$) ratio gradients and observational/seeing 
effects, the prediction is trivial. Neither of these effects has any affect on either 
the galaxy stellar mass or the BH mass -- systems simply evolve passively 
and appear to change in radius. 

The case of stellar mass uncertainties is only slightly more complex. Here, 
there is no effect that can change the BH mass; but at high redshifts, the 
system {\em appears} to be higher-stellar mass, by an amount tuned to match the 
apparent evolution in the size-mass relation. Therefore, for an (apparent) mass-selected sample 
at high redshift, since the ``true'' stellar mass is lower, the BH mass must be lower, 
and there will appear to be strong negative evolution in $M_{\rm BH}$. 

Adiabatic expansion requires significant host bulge mass loss to have occurred since the 
time of BH and bulge formation, so it predicts that BH masses
at high redshifts should be substantially smaller than those today 
(at fixed stellar mass; such that after this host mass loss they will lie on the 
$z=0$ BH-host relation). In the prediction shown in Figure~\ref{fig:bhs} ({\em top}), 
we assume that the BH does not grow in 
mass over this time (i.e.\ the entire $z=0$ BH mass is in place at the time of 
formation). It is possible that, instead, some small fraction of this material from adiabatic mass 
loss is consumed by the BH, increasing its mass. 
However, this means that the BH mass at the time of formation must be lower; 
therefore, the evolution in $M_{\rm BH}$ at fixed bulge mass would be even {\em more} 
negative (towards lower $M_{\rm BH}(z)/M_{\rm BH}(0)$). Likewise for the other models 
above, if further accretion occurs subsequent to the initial bulge formation. 

The case of identical dry mergers is also straightforward. Unlike the cases above, 
the BH will grow in time -- with each dry merger, 
the BH should grow via the merger of the two progenitor BHs. But since 
the systems are, by definition, structurally identical, they should obey the same 
proportionality between BH mass and host galaxy stellar mass. So BH mass 
and stellar mass simply add linearly, and the BH-host mass relation (being linear itself) is conserved 
by these mergers. As a consequence, there is no predicted evolution 
in $M_{\rm BH}$ at fixed stellar mass (i.e.\ $M_{\rm BH}/M_{\rm bulge}$), 
although $M_{\rm BH}$ and $M_{\rm bulge}$ can both grow significantly 
for an individual system. 

The minor/late merger case is more interesting. 
As before, BH mass will grow via mergers. However, unlike the identical dry 
merger case, it is not necessarily true that the BH masses of both merging 
systems obey the same BH-host galaxy correlations (and therefore that 
BH and host mass will add linearly and conserve the $M_{\rm BH}-M_{\rm bulge}$ 
relation). The previous scale-invariance can be broken in two ways. First, imagine 
the case of a merger with a late-forming, gas-poor disk-dominated 
galaxy (for the case of simplicity, take the extreme limit of a gas-free, 
nearly pure disk secondary). Since the secondary has little or no gas, there will 
be no new accretion; since it has little or no bulge, it has corresponding 
little or no BH. As a consequence, the secondary merger contributes negligible BH mass. 
But the entire secondary disk stellar mass will be violently relaxed, adding 
substantially to the total bulge mass. 
Similarly, spheroids and disks that merge later will have formed later, 
from more gas-poor mergers. Such mergers build up a less-deep central potential 
and fuel less material to the BH, and so will form less massive BHs, relative to their 
bulge masses \citep[see e.g.][]{hopkins:bhfp.theory}. 
The result in either case is that $M_{\rm BH}$ at high redshifts should be 
{\em larger} at fixed stellar mass. 

An upper limit to the magnitude of this effect is easy to determine. In this case, assume 
that no BH mass is contributed from the subsequent late/minor mergers. 
This could be the case if e.g.\ the depletion of gas fractions occurs sufficiently 
rapidly with cosmic time, or if the late/minor mergers involve preferentially very disk-dominated 
galaxies, or if BH-BH mergers are inefficient and the systems are relatively gas-poor. 
Because the BH mass still cannot decrease, and must lie on the observed $z=0$ 
correlations, this amounts to assuming the entire $z=0$ BH mass is in place at 
the time of formation, with only bulge mass added since. The resulting prediction is 
up to a factor $\sim 3$ increase in $M_{\rm BH}/M_{\rm bulge}$ from $z=0-2$. 

A more realistic prediction requires some model for how massive a BH will be as a function of 
bulge properties at formation (e.g.\ gas richness, velocity dispersion, mass, etc.), to determine 
how much BH mass is contributed by the late/secondary mergers. 
One such model is presented in detail in \citet{hopkins:cusps.ell}; similar 
cases are discussed in \citet{hopkins:bhfp.theory} and \citet{croton:msigma.evolution}. It is not 
our intention here to develop those models (their predictions for the evolution in BH-host 
correlations are discussed in much greater detail therein); but we briefly summarize the 
salient points and show the results. 

As in most models, BH growth is regulated by some form of feedback, from 
energy or momentum associated with the accretion luminosity. 
Various works have shown that this can naturally account for the 
observed correlations between BH mass and spheroid velocity dispersion, 
mass, and other properties \citep[see e.g.][]{silkrees:msigma,dimatteo:msigma,
hopkins:seyferts,hopkins:bhfp.theory}.
Generically, in such models, the ``most fundamental'' correlation is between 
BH mass and some measure of the binding energy of the material which must 
be ``halted'' to arrest further accretion; \citet{hopkins:bhfp.theory} show, 
in high-resolution hydrodynamic simulations, that this can be approximated 
by a ``true'' fundamental plane-like correlation between BH mass and 
the combination of stellar mass and $\sigma$, $M_{\rm BH}\propto M_{\ast}^{0.5}\,\sigma^{2.2}$, 
{\em at the time of bulge formation} in gas rich mergers. 
Direct analysis of the observed BH-host correlations appear to support such a 
driving correlation 
\citep[see e.g.][]{hopkins:bhfp.obs,aller:mbh.esph}. Moreover, \citet{younger:minor.mergers} 
showed that such a model 
predicts a unique difference between e.g.\ the $M_{\rm BH}-\sigma$ relation of 
``classical'' bulges (believed to be formed in mergers) and ``pseudobulges'' (formed 
in secular events, from e.g.\ disk bars), because the observed velocity dispersion relates 
differently to the central binding energy in structurally distinct objects. Since then, 
various observations have found that such a difference exists, with the sense and 
magnitude predicted \citep{hu:msigma.pseudobulges,
greene:pseudobulge.msigma,gadotti:pseudobulge.msigma}. 
In any case given such a model for the BH mass as a function of bulge properties 
(at e.g.\ fixed mass or $\sigma$), it is straightforward to calculate the contribution 
from subsequent minor/late mergers (which we do 
following, in detail, the equations and approach in \citet{hopkins:cusps.ell}) 
and (given the $z=0$ constraint) predicted 
evolution in Figure~\ref{fig:bhs}. 

Essentially, the result is similar, but with somewhat weaker evolution than the 
upper limit case (some non-trivial BH mass is contributed by these subsequent mergers). 
The net evolution, as discussed in 
\citet{hopkins:bhfp.theory} and \citet{hopkins:cusps.evol} 
owes to the fact that because 
high redshift mergers are more gas rich and yield more compact remnants, 
they have higher central binding energies than low-redshift counterparts 
(for the same stellar mass) and therefore will produce higher-mass BHs. 
The bulge mass ``catches up'' via accretion of less dense, later-forming systems 
with lower BH mass for their bulge, and (in some cases) less bulge, 
as outlined in \citet{croton:msigma.evolution}. 

We compare these predictions to various observational estimates of the 
evolution. Specifically, we consider observations from 
$z=0-1$ from SDSS AGN samples \citep{salviander:msigma.evolution,
salviander:midz.msigma.evol}, from $z=1-3$ from 
lensed quasar hosts \citep{peng:magorrian.evolution}, 
and at $z=0.36$ from narrow-redshift Seyfert 
samples \citep{woo06:lowz.msigma.evolution,treu:msigma.evol}. 
We caution, however, that these estimates (indirect BH mass estimates based on the virial 
mass indicators) are systematically uncertain, and various biases are present 
in AGN-selected samples that will tend to systematically over-estimate the amount of 
evolution at high redshift \citep[see][]{lauer:mbh.bias} \citep[although the][estimates do 
attempt to correct for this bias]{salviander:midz.msigma.evol}. 
Alternatively, \citet{hopkins:msigma.limit} determine a non-parametric upper limit to the 
degree of evolution by comparing the observed spheroid mass density at each redshift to the 
BH mass density at $z=0$, given the constraint that BHs cannot decrease in mass; we show 
this as well. Other indirect constraints, from e.g.\ clustering and integral constraints, 
give qualitatively similar results, but with a large scatter 
\citep[see e.g.][]{merloni:magorrian.evolution,mclure.dunlop:mbh,
alexander:bh.growth,alexander:smg.bh.masses,
adelbergersteidel:magorrian.evolution,fine:mbh-mhalo.clustering}. 
Although the various biases and uncertainties involved are large, it appears robust that 
there is probably {\em some} evolution towards higher $M_{\rm BH}$ at 
fixed galaxy stellar mass. We hesitate to use this as a quantitative 
constraint on the models here, but note that this appears to strongly conflict with the 
adiabatic expansion and stellar mass error models, and moreover stress that 
{\em only} the late/minor merger channel provides a mean to evolve in the 
observed sense, at any level.

\section{Discussion}
\label{sec:discuss}

We have illustrated how different physical effects combine to explain 
the observationally inferred size evolution of massive ellipticals with redshift, 
and shown how various observations can distinguish between these models. 

Observations have shown that, at each time, most of the spheroid population 
is recently assembled. As a consequence, the evolution in the size-mass 
relation {\em as a whole} must reflect a redshift dependent scaling in 
the in-situ sizes of ellipticals {\em at the time of formation}. 
As is discussed in detail in \citet{khochfar:size.evolution.model} and 
\citet{hopkins:cusps.evol}, 
this occurs naturally in merger models: at the same mass, higher-redshift 
disks have larger observed gas fractions. The gas fraction at the time of 
merger, which yields the compact, remnant starburst stellar populations 
required to explain elliptical stellar population gradients, kinematics, 
mass profile shapes, and densities, is the dominant determinant of the 
size of any merger remnant \citep[see e.g.][and references therein]{hopkins:cusps.fp}.
As a consequence, 
mergers of these gas-rich disks at high 
redshifts will yield spheroids with smaller effective radii. 
We show that simulated merger remnants with the appropriate gas fractions 
as a function of redshift naturally reproduce the observed evolution in the 
mean trends. 

That being said, there does not appear to be a ``relic'' population of 
high-mass, very compact spheroids \citep[see e.g.][]{trujillo:dense.gal.nearby}. 
It may even be the case that the most highly clustered (and therefore 
earliest-to-assemble) spheroids (i.e.\ BCGs) may have somewhat larger 
sizes for their mass. 
Therefore, some other process must increase the apparent sizes of these systems 
after their formation, at least keeping pace with 
the evolution in the median population owing to redshift-dependent 
gas fractions. 

We discuss six effects that can contribute to such evolution, that cover the 
range of possibilities in the literature. 
These include: ``identical'' dry mergers (mergers of identically compact spheroids), 
``late/minor accretion (major {\em or} minor mergers at later times, with less 
dense material -- for example, less gas-rich disks or other spheroids formed in 
lower-redshift, less gas-rich mergers), adiabatic expansion 
(owing to mass expulsion from 
feedback or gradual stellar mass loss from stellar population 
evolution), the presence and subsequent evolution of stellar mass-to-light 
ratio gradients (changing e.g.\ the observed optical or near-IR $R_{e}$ 
relative to the stellar mass $R_{e}$), 
age or redshift-dependent uncertainties in stellar masses 
(owing to e.g.\ unaccounted-for contributions of AGN star light in young 
stellar populations), and seeing/observational effects (e.g.\ possible 
redshift-dependent fitting biases from limited dynamic range and 
resolution effects, or issues arising from different choices of definition of 
effective radii and different distributions of galaxy shapes in different samples). 

In each case, we show how these different mechanisms should 
move systems through the size-mass space with redshift, and 
how this relates to other observable properties, including spheroid 
velocity dispersions, central stellar mass densities, light profile shapes, 
and black hole-host galaxy correlations. 

Given this, we construct the predictions from each model, if it were solely 
responsible for the observed evolution. 
In particular, we demand that each model obey the observational constraints 
as boundary conditions: newly-formed 
ellipticals must appear to have the appropriate observed size at each redshift, 
and then, evolved forward allowing only the given affect to alter the size, 
must lie on the appropriate parameter correlations of $z=0$ galaxies 
(i.e.\ have velocity dispersions, densities, and profile 
shapes at $z=0$ within the range observed for systems of the remnant final mass). 
This allows us to construct the corresponding predictions from 
each model in these other observable properties, 
and in particular in how these properties should scale with fixed 
mass as a function of redshift. 

Comparing these to recent observational constraints, 
we can clearly rule out several of the above models as a dominant 
driver of size evolution. The {\em only} model 
that is consistent with all of these constraints is the ``late/minor merger'' 
model. \citet{hopkins:density.galcores} have shown, for example, 
that the observed profiles of the high-redshift systems are 
actually very similar to the observed cores of massive ellipticals today -- 
they are not more dense, by even a factor $\sim$a few 
\citep[see also][who reach similar conclusions]{bezanson:massive.gal.cores.evol}. 
It appears that the dominant effect has been the buildup of extended, 
lower-density wings, from high redshifts to today, yielding higher 
Sersic indices in low-redshift systems and more extended envelopes 
that drive the {\rm effective} radii to larger values, while 
leaving the central properties relatively intact. 
Supporting evidence for this comes form \citet{cenarro:sigma.of.highz.compact.ell}, 
who see only weak growth in the velocity dispersions of 
galaxies at fixed mass with redshift, consistent with evolution via buildup of 
low-density material that does not strongly affect $\sigma$. 
In contrast, an adiabatic contraction or identical dry merger model 
would predict dramatically higher central densities at high redshifts, 
with correspondingly larger velocity dispersions. Invoking 
stellar mass biases would predict dramatic inverse evolution in velocity 
dispersions with redshift. And invoking mass-to-light ratio gradients 
predicts weaker velocity dispersion evolution, and opposite light 
profile shape evolution, relative to that observed. 

However, although it is in principle possible to explain the entire necessary 
size evolution via ``late/minor mergers'' 
without violating these observational constraints, both theoretical models 
and independent constraints on e.g.\ merger rates and the stellar 
mass function suggest that the required number of mergers would be 
too large, at least in the extreme case of factor $\sim6-10$ evolution in sizes 
of the most massive, early-forming spheroids as required. 
A priori theoretical models and semi-empirical models based on observed 
clustering and stellar mass function evolution predict closer to 
a factor $\sim3-4$ size evolution via this channel, given the 
observed/expected merger rates. 

We therefore consider a high-resolution case study, using 
hydrodynamic merger simulations, of a typical history of such an 
early-forming galaxy in one such cosmological model. 
We show that, as required by the observations, the minor/late 
merger channel does dominate the size evolution, yielding a factor 
$\sim3-4$ size evolution with a moderate (factor $\sim1.5-2$) stellar mass
growth. However, the other effects described here also all contribute 
some apparent or real size evolution. Each of the them
contributes a relatively small effect, $\sim20-30\%$. 
Mass-to-light ratio gradients are inevitably present at all times, 
contributing a net effect at this level via slight bias towards smaller 
$B$-band $R_{e}$ at early times and larger $R_{e}$ at late times, but 
each at the $\sim10\%$ level. Observational fitting/resolution effects, 
and/or small shifts in the typical flattening of systems in observed samples 
coupled to the adoption of a ``circularized'' effective radius definition, 
can contribute a relatively small $\sim20\%$ to size estimates, 
for the observed and model profile shapes. Given the stellar 
ages, star formation rates, and masses already in place at high redshift, and low-redshift 
constraints on star formation histories of massive ellipticals, 
there could not be a very large gas supply awaiting removal in the 
compact passive systems, but they will inevitably lose $\sim20\%$ of their 
mass to later stellar evolution, and will adiabatically expand as a result. 
Also, given the observed photometry, stellar mass estimates appear uncertain 
at less than the $\sim20\%$ level. 
And the earliest (highest redshift) mergers after the system first forms are 
likely to be similarly gas-rich and/or compact, contributing an average of 
$\sim20\%$ in mass via the ``identical'' dry (or mixed) mergers channel. 

Each of these effects is small; together, however, they contribute a 
net additional factor $\sim2-3$ size evolution. This is the remaining 
factor needed to explain the observed trends. The predictions for other quantities 
such as velocity dispersions, central densities, and profile shapes, 
in such a realistic model, are still very similar to the minor/late merger 
track predictions (not surprising, given that this effect still dominates), 
and as such consistent with observations. But with a net mass evolution 
of just a factor $\sim2-3$, consistent with merger rates and stellar mass 
function evolution, the system will expand by nearly an order of magnitude in  
apparent $R_{e}$. The combination of secondary effects -- all similarly 
important -- explains the remaining factor needed to reconcile 
models of e.g.\ mergers alone and the observations. Moreover, 
because each of these effects is, individually, small, 
the success of theoretical models does not critically depend on any one 
of the effects operating (other than, of course, late/minor mergers). 
The size evolution of even the most massive, early 
forming spheroids, therefore, appears to be a natural consequence of 
hierarchical, merger-driven spheroid formation; but understanding this 
evolution in detail requires accounting for a number of
effects related to e.g.\ stellar populations, small deviations in profile shape, 
multi-component stellar mass profiles, and different mergers 
that can only be followed in high-resolution simulations. 

Interestingly, it also appears that only the ``late/minor merger'' channel 
provides a means for evolution in the BH-host galaxy correlations, in the 
sense of more massive BHs at high redshift relative to their host galaxies. 
The other mechanisms predict either no or inverse evolution 
in these correlations, which both appear to conflict with the observations and 
would make reconciling the BH population and high-redshift quasar population 
(which requires a large fraction of the present-day massive 
BH population be extant and active 
at these times \citep{hopkins:bol.qlf}) difficult. The full model above 
appears to be consistent with recent observations of such 
evolution, but observational uncertainties remain large. In this scenario, 
early-forming BHs are relatively massive, owing to the deep potential 
wells of their compact hosts; the spheroid mass can then ``catch up'' to the $z=0$ 
relation observed via accretion of lower density material -- e.g.\ gas-poor disks or 
spheroids formed from mergers of such disks, with relatively small pre-existing BH masses. 
For more details, we refer to \citet{hopkins:cusps.ell}. 

Further observational tests will present valuable constraints of these models. 
We have outlined predictions for a number of basic properties, 
for which observations have only just become possible at intermediate 
redshifts. Better observations as a function of mass, redshift, and galaxy 
morphology remain critical. The present velocity dispersion, central 
stellar density, and profile shape information is limited to samples of 
a few objects and, often, stacked images. 
Direct construction of surface density profiles, 
in order to measure the mass at low stellar densities as a function of 
redshift, can map out the effects of less dense later mergers. 
Imaging in multiple bands (yielding e.g.\ color gradients) 
can greatly constrain the role of stellar 
mass-to-light ratio gradients and stellar mass biases. Better constraints 
on stellar populations at low redshift -- in particular stellar population age 
gradients -- will constrain how efficient stellar mass loss must be, 
and correspondingly the role of adiabatic expansion. 
Kinematics of galaxies at all redshifts represent a powerful constraint 
both on the {\em original}, gas-rich merger history (allowing 
direct tests at high redshift of the hypothesis, which appears successful 
at low redshift, that gas-richness dominates the determination of the 
remnant size) and on the subsequent merger history, 
potentially mapping between gas-rich or equal density mergers 
and dry, lower-density mergers \citep{mihos:cusps,
barneshernquist96,naab:gas,
cox:kinematics,robertson:fp,onorbe:diss.fp.details,
jesseit:kinematics,hopkins:cusps.ell,hopkins:cores}. 

We wish to emphasize that the various observations discussed here, 
and the magnitude or nature of size evolution, do {\em not} 
themselves strongly constrain whether or not the ``late/minor accretion'' channel 
is actually dominated by a few ``major'' or a larger number of ``minor'' mergers. 
Provided that the same {\em total} amount of low-density 
material (stellar mass) is added to a system, in a 
dissipationless/collisionless fashion, it makes no difference whether or not 
it is ``brought in'' by a major or minor merger. In other words, so long as the major companions 
are sufficiently low-density (unlike in the ``identical dry mergers'' case, where they 
are, by definition, high-density), the characteristic stellar densities and 
radii of that material in the post-merger remnant will be the {\em same} 
as if the material was merged via a series of much more minor mergers. 
This follows necessarily from basic phase-space considerations 
\citep[e.g.][]{gallagherostriker72,
hernquist:phasespace}, 
and has been seen as well in various numerical studies 
\citep{bournaud:minor.mergers,
boylankolchin:dry.mergers,hopkins:cores,
naab:size.evol.from.minor.mergers,feldmann:bgg.size.evol.in.hydro.sims}. 
Realistic cosmological models predict that both major and 
minor mergers contribute comparably to this aggregation of low-density 
material \citep[see e.g.][and references therein]{hopkins:merger.rates}, 
with major mergers dominating at masses near and somewhat above 
$\sim L_{\ast}$, and minor mergers becoming progressively more important 
at higher masses (essentially, most of the mass is near $\sim L_{\ast}$; 
so a ``typical'' $z=0$, $\sim 10\,L_{\ast}$ system will reflect buildup from 
$\sim$1:10 mergers). 

There will of course be considerable galaxy-to-galaxy variation in 
merger histories, and some fraction $\sim10\%$ of systems may 
escape much subsequent merging \citep{hopkins:cusps.evol}, 
suggestively similar to the recently-observed abundance of 
populations of present-day compact systems in galaxy clusters, 
their expected environments \citep{valentinuzzi:superdense.local.ell.wings}. 
Distinguishing observationally between 
e.g.\ a series of minor mergers 
and a couple of major mergers will require additional, independent checks. 
Obviously, direct constraints on merger rates/fractions as a function of 
mass, redshift, and mass ratio are important. Also, 
higher-order kinematics, such as e.g.\ galaxy isotropies 
and the presence/absence of certain kinematic subsystems 
can distinguish between different merger histories 
\citep[see references above and e.g.][]{burkert:anisotropy,
hoffman:dissipation.and.gal.kinematics}. 
These constraints will be particularly 
valuable in informing theoretical models that attempt to follow the detailed 
formation histories of such systems. 

However, it should also be emphasized, as 
above, that the vast majority of ellipticals/spheroids 
do {\em not} form via these high redshift, compact channels. 
The observed mass density of bulge-dominated galaxies at $z\sim2$ is 
only $\sim5\%$ of its $z=0$ value \citep[and this is probably an upper 
limit, given the expected contamination in how such systems are 
selected; see e.g.][]{daddi05:drgs,labbe05:drgs,grazian:drg.comparisons}; 
by $z=1$ it has increased but is still only $\sim20-35\%$ of 
its $z=0$ value \citep{bundy:mfs,bundy:mtrans,
abraham:red.mass.density}. So most bulges are formed at relatively late times, 
where they are not observed to be especially compact, and require no subsequent 
merging or other activity to change their profile shapes/sizes. 
This is particularly important for the ``cusp'' elliptical population, 
of rapidly rotating, diskier ellipticals and S0's, which are believed to be the direct 
products of gas-rich mergers \citep[and cannot ``tolerate'' much subsequent dry merging 
without disrupting these properties; see e.g.][]{naab:dry.mergers,
ciotti:dry.vs.wet.mergers,hopkins:cores,cox:remerger.kinematics.prep}; these dominate the 
bulge/spheroid/elliptical mass density, and dominate the population by number 
at all but the highest masses $>10^{11}\,\msun$, the regime where 
``core'' ellipticals (believed to have survived dry mergers) become important, 
in good agreement with the models for evolution considered here.

\acknowledgments 
We thank Eliot Quataert, Norm Murray, Pieter van Dokkum, and 
Ignacio Trujillo for a number of very helpful conversations and suggestions 
throughout the development of this manuscript. 
Support for PFH was provided by the Miller Institute for Basic Research 
in Science, University of California Berkeley. SW and TJC gratefully 
acknowledge support from the W.~M.\ Keck Foundation.
\\

\bibliography{/Users/phopkins/Documents/lars_galaxies/papers/ms}

\begin{thebibliography}{178}
\expandafter\ifx\csname natexlab\endcsname\relax\def\natexlab#1{#1}\fi

\bibitem[{{Abraham} {et~al.}(2007)}]{abraham:red.mass.density}
{Abraham}, R.~G., {et~al.} 2007, \apj, 669, 184

\bibitem[{{Adelberger} \&
  {Steidel}(2005)}]{adelbergersteidel:magorrian.evolution}
{Adelberger}, K.~L., \& {Steidel}, C.~C. 2005, \apjl, 627, L1

\bibitem[{{Akiyama} {et~al.}(2008){Akiyama}, {Minowa}, {Kobayashi}, {Ohta},
  {Ando}, \& {Iwata}}]{akiyama:lbg.weak.size.evol}
{Akiyama}, M., {Minowa}, Y., {Kobayashi}, N., {Ohta}, K., {Ando}, M., \&
  {Iwata}, I. 2008, \apjs, 175, 1

\bibitem[{{Alexander} {et~al.}(2005){Alexander}, {Smail}, {Bauer}, {Chapman},
  {Blain}, {Brandt}, \& {Ivison}}]{alexander:bh.growth}
{Alexander}, D.~M., {Smail}, I., {Bauer}, F.~E., {Chapman}, S.~C., {Blain},
  A.~W., {Brandt}, W.~N., \& {Ivison}, R.~J. 2005, \nat, 434, 738

\bibitem[{{Alexander} {et~al.}(2008)}]{alexander:smg.bh.masses}
{Alexander}, D.~M., {et~al.} 2008, \aj, 135, 1968

\bibitem[{{Aller} \& {Richstone}(2007)}]{aller:mbh.esph}
{Aller}, M.~C., \& {Richstone}, D.~O. 2007, \apj, 665, 120

\bibitem[{{Barden} {et~al.}(2005)}]{barden:disk.size.evol}
{Barden}, M., {et~al.} 2005, \apj, 635, 959

\bibitem[{{Barnes} \& {Hernquist}(1992)}]{barneshernquist92}
{Barnes}, J.~E., \& {Hernquist}, L. 1992, \araa, 30, 705

\bibitem[{{Barnes} \& {Hernquist}(1996)}]{barneshernquist96}
---. 1996, \apj, 471, 115

\bibitem[{{Barnes} \& {Hernquist}(1991)}]{barnes.hernquist.91}
{Barnes}, J.~E., \& {Hernquist}, L.~E. 1991, \apjl, 370, L65

\bibitem[{{Batcheldor} {et~al.}(2007){Batcheldor}, {Marconi}, {Merritt}, \&
  {Axon}}]{batcheldor:bcgs}
{Batcheldor}, D., {Marconi}, A., {Merritt}, D., \& {Axon}, D.~J. 2007, \apjl,
  663, L85

\bibitem[{{Bell} \& {de Jong}(2001)}]{belldejong:tf}
{Bell}, E.~F., \& {de Jong}, R.~S. 2001, \apj, 550, 212

\bibitem[{{Bell} {et~al.}(2003){Bell}, {McIntosh}, {Katz}, \&
  {Weinberg}}]{bell:mfs}
{Bell}, E.~F., {McIntosh}, D.~H., {Katz}, N., \& {Weinberg}, M.~D. 2003, \apjs,
  149, 289

\bibitem[{{Bell} {et~al.}(2006)}]{bell:dry.mergers}
{Bell}, E.~F., {et~al.} 2006, \apj, 640, 241

\bibitem[{{Bernardi} {et~al.}(2007){Bernardi}, {Hyde}, {Sheth}, {Miller}, \&
  {Nichol}}]{bernardi:bcg.scalings}
{Bernardi}, M., {Hyde}, J.~B., {Sheth}, R.~K., {Miller}, C.~J., \& {Nichol},
  R.~C. 2007, \aj, 133, 1741

\bibitem[{{Bezanson} {et~al.}(2009){Bezanson}, {van Dokkum}, {Tal},
  {Marchesini}, {Kriek}, {Franx}, \& {Coppi}}]{bezanson:massive.gal.cores.evol}
{Bezanson}, R., {van Dokkum}, P.~G., {Tal}, T., {Marchesini}, D., {Kriek}, M.,
  {Franx}, M., \& {Coppi}, P. 2009, \apj, 697, 1290

\bibitem[{{Borch} {et~al.}(2006)}]{borch:mfs}
{Borch}, A., {et~al.} 2006, \aap, 453, 869

\bibitem[{{Bournaud} {et~al.}(2005){Bournaud}, {Jog}, \&
  {Combes}}]{bournaud:minor.mergers}
{Bournaud}, F., {Jog}, C.~J., \& {Combes}, F. 2005, \aap, 437, 69

\bibitem[{{Boylan-Kolchin} {et~al.}(2005){Boylan-Kolchin}, {Ma}, \&
  {Quataert}}]{boylankolchin:mergers.fp}
{Boylan-Kolchin}, M., {Ma}, C.-P., \& {Quataert}, E. 2005, \mnras, 362, 184

\bibitem[{{Boylan-Kolchin} {et~al.}(2006){Boylan-Kolchin}, {Ma}, \&
  {Quataert}}]{boylankolchin:dry.mergers}
---. 2006, \mnras, 369, 1081

\bibitem[{{Bridge} {et~al.}(2009)}]{bridge:merger.fraction.new.prep}
{Bridge}, C.~R., {et~al.} 2009, \apj, in preparation

\bibitem[{{Brown} {et~al.}(2007){Brown}, {Dey}, {Jannuzi}, {Brand}, {Benson},
  {Brodwin}, {Croton}, \& {Eisenhardt}}]{brown:mf.evolution}
{Brown}, M.~J.~I., {Dey}, A., {Jannuzi}, B.~T., {Brand}, K., {Benson}, A.~J.,
  {Brodwin}, M., {Croton}, D.~J., \& {Eisenhardt}, P.~R. 2007, \apj, 654, 858

\bibitem[{{Bruzual} \& {Charlot}(2003)}]{BC03}
{Bruzual}, G., \& {Charlot}, S. 2003, \mnras, 344, 1000

\bibitem[{{Buitrago} {et~al.}(2008){Buitrago}, {Trujillo}, {Conselice},
  {Bouwens}, {Dickinson}, \& {Yan}}]{buitrago:highz.size.evol}
{Buitrago}, F., {Trujillo}, I., {Conselice}, C.~J., {Bouwens}, R.~J.,
  {Dickinson}, M., \& {Yan}, H. 2008, \apjl, 687, L61

\bibitem[{{Bundy} {et~al.}(2005){Bundy}, {Ellis}, \& {Conselice}}]{bundy:mfs}
{Bundy}, K., {Ellis}, R.~S., \& {Conselice}, C.~J. 2005, \apj, 625, 621

\bibitem[{{Bundy} {et~al.}(2006)}]{bundy:mtrans}
{Bundy}, K., {et~al.} 2006, \apj, 651, 120

\bibitem[{{Burkert} {et~al.}(2008){Burkert}, {Naab}, {Johansson}, \&
  {Jesseit}}]{burkert:anisotropy}
{Burkert}, A., {Naab}, T., {Johansson}, P.~H., \& {Jesseit}, R. 2008, \apj,
  685, 897

\bibitem[{{Calura} {et~al.}(2008){Calura}, {Jimenez}, {Panter}, {Matteucci}, \&
  {Heavens}}]{calura:sdss.gas.fracs}
{Calura}, F., {Jimenez}, R., {Panter}, B., {Matteucci}, F., \& {Heavens}, A.~F.
  2008, \apj, 682, 252

\bibitem[{{Caon} {et~al.}(1990){Caon}, {Capaccioli}, \& {Rampazzo}}]{caon90}
{Caon}, N., {Capaccioli}, M., \& {Rampazzo}, R. 1990, \aaps, 86, 429

\bibitem[{{Cappellari} {et~al.}(2009){Cappellari}, {di Serego Alighieri},
  {Cimatti}, {Daddi}, {Renzini}, {Kurk}, {Cassata}, {Dickinson},
  {Franceschini}, {Mignoli}, {Pozzetti}, {Rodighiero}, {Rosati}, \&
  {Zamorani}}]{cappellari:highz.gal.sigmas.and.mdyn}
{Cappellari}, M., {di Serego Alighieri}, S., {Cimatti}, A., {Daddi}, E.,
  {Renzini}, A., {Kurk}, J.~D., {Cassata}, P., {Dickinson}, M., {Franceschini},
  A., {Mignoli}, M., {Pozzetti}, L., {Rodighiero}, G., {Rosati}, P., \&
  {Zamorani}, G. 2009, \apjl, in press, arXiv:0906.3648

\bibitem[{{Cenarro} \& {Trujillo}(2009)}]{cenarro:sigma.of.highz.compact.ell}
{Cenarro}, A.~J., \& {Trujillo}, I. 2009, \apjl, 696, L43

\bibitem[{{Chabrier}(2003)}]{chabrier:imf}
{Chabrier}, G. 2003, \pasp, 115, 763

\bibitem[{{Cimatti} {et~al.}(2008)}]{cimatti:highz.compact.gal.smgs}
{Cimatti}, A., {et~al.} 2008, \aap, 482, 21

\bibitem[{{Ciotti} {et~al.}(2007){Ciotti}, {Lanzoni}, \&
  {Volonteri}}]{ciotti:dry.vs.wet.mergers}
{Ciotti}, L., {Lanzoni}, B., \& {Volonteri}, M. 2007, \apj, 658, 65

\bibitem[{{Ciotti} \& {Ostriker}(2007)}]{ciottiostriker:recycling}
{Ciotti}, L., \& {Ostriker}, J.~P. 2007, \apj, 665, 1038

\bibitem[{{Covington} {et~al.}(2008){Covington}, {Dekel}, {Cox}, {Jonsson}, \&
  {Primack}}]{covington:diss.size.expectation}
{Covington}, M., {Dekel}, A., {Cox}, T.~J., {Jonsson}, P., \& {Primack}, J.~R.
  2008, \mnras, 384, 94

\bibitem[{{Cox} {et~al.}(2009)}]{cox:remerger.kinematics.prep}
{Cox}, T., {et~al.} 2009, \apj, in preparation

\bibitem[{{Cox} {et~al.}(2006{\natexlab{a}}){Cox}, {Di Matteo}, {Hernquist},
  {Hopkins}, {Robertson}, \& {Springel}}]{cox:xray.gas}
{Cox}, T.~J., {Di Matteo}, T., {Hernquist}, L., {Hopkins}, P.~F., {Robertson},
  B., \& {Springel}, V. 2006{\natexlab{a}}, \apj, 643, 692

\bibitem[{{Cox} {et~al.}(2006{\natexlab{b}}){Cox}, {Dutta}, {Di Matteo},
  {Hernquist}, {Hopkins}, {Robertson}, \& {Springel}}]{cox:kinematics}
{Cox}, T.~J., {Dutta}, S.~N., {Di Matteo}, T., {Hernquist}, L., {Hopkins},
  P.~F., {Robertson}, B., \& {Springel}, V. 2006{\natexlab{b}}, \apj, 650, 791

\bibitem[{{Cresci} {et~al.}(2009)}]{cresci:dynamics.highz.disks}
{Cresci}, G., {et~al.} 2009, \apj, 697, 115

\bibitem[{{Croton}(2006)}]{croton:msigma.evolution}
{Croton}, D.~J. 2006, \mnras, 369, 1808

\bibitem[{{Daddi} {et~al.}(2005)}]{daddi05:drgs}
{Daddi}, E., {et~al.} 2005, \apj, 626, 680

\bibitem[{{Damjanov} {et~al.}(2009)}]{damjanov:red.nuggets}
{Damjanov}, I., {et~al.} 2009, \apj, 695, 101

\bibitem[{{Darg}
  {et~al.}(2009{\natexlab{a}})}]{darg:galzoo.merger.frac.by.morph}
{Darg}, D.~W., {et~al.} 2009{\natexlab{a}}, \mnras, in press, arXiv:0903.4937

\bibitem[{{Darg} {et~al.}(2009{\natexlab{b}})}]{darg:galzoo.merger.properties}
---. 2009{\natexlab{b}}, \mnras, in press, arXiv:0903.5057

\bibitem[{{Dav{\'e}}(2008)}]{dave:imf.evol}
{Dav{\'e}}, R. 2008, \mnras, 385, 147

\bibitem[{{Dehnen}(1993)}]{dehnen:1993.hernquist.profile.families}
{Dehnen}, W. 1993, \mnras, 265, 250

\bibitem[{{Di Matteo} {et~al.}(2005){Di Matteo}, {Springel}, \&
  {Hernquist}}]{dimatteo:msigma}
{Di Matteo}, T., {Springel}, V., \& {Hernquist}, L. 2005, \nat, 433, 604

\bibitem[{{Erb}(2008)}]{erb:outflow.inflow.masses}
{Erb}, D.~K. 2008, \apj, 674, 151

\bibitem[{{Erb} {et~al.}(2006){Erb}, {Steidel}, {Shapley}, {Pettini}, {Reddy},
  \& {Adelberger}}]{erb:lbg.gasmasses}
{Erb}, D.~K., {Steidel}, C.~C., {Shapley}, A.~E., {Pettini}, M., {Reddy},
  N.~A., \& {Adelberger}, K.~L. 2006, \apj, 646, 107

\bibitem[{{Fan} {et~al.}(2008){Fan}, {Lapi}, {De Zotti}, \&
  {Danese}}]{fan:adiabatic.expansion.ell.size.evol}
{Fan}, L., {Lapi}, A., {De Zotti}, G., \& {Danese}, L. 2008, \apjl, 689, L101

\bibitem[{{Feldmann} {et~al.}(2009){Feldmann}, {Carollo}, {Mayer}, {Renzini},
  {Lake}, {Quinn}, {Stinson}, \&
  {Yepes}}]{feldmann:bgg.size.evol.in.hydro.sims}
{Feldmann}, R., {Carollo}, C.~M., {Mayer}, L., {Renzini}, A., {Lake}, G.,
  {Quinn}, T., {Stinson}, G.~S., \& {Yepes}, G. 2009, \apj, in press,
  arXiv:0906.3022

\bibitem[{{Ferguson} {et~al.}(2004)}]{ferguson:disk.size.evol}
{Ferguson}, H.~C., {et~al.} 2004, \apjl, 600, L107

\bibitem[{{Ferrarese} {et~al.}(2006)}]{ferrarese:profiles}
{Ferrarese}, L., {et~al.} 2006, \apjs, 164, 334

\bibitem[{{Fine} {et~al.}(2006)}]{fine:mbh-mhalo.clustering}
{Fine}, S., {et~al.} 2006, \mnras, 373, 613

\bibitem[{{Forster Schreiber}
  {et~al.}(2009)}]{forsterschreiber:z2.sf.gal.spectroscopy}
{Forster Schreiber}, N.~M., {et~al.} 2009, \apj, in press, arXiv:0903.1872

\bibitem[{{Franceschini} {et~al.}(2006)}]{franceschini:mfs}
{Franceschini}, A., {et~al.} 2006, \aap, 453, 397

\bibitem[{{Franx} {et~al.}(2008){Franx}, {van Dokkum}, {Schreiber}, {Wuyts},
  {Labb{\'e}}, \& {Toft}}]{franx:size.evol}
{Franx}, M., {van Dokkum}, P.~G., {Schreiber}, N.~M.~F., {Wuyts}, S.,
  {Labb{\'e}}, I., \& {Toft}, S. 2008, \apj, 688, 770

\bibitem[{{Gadotti} \& {Kauffmann}(2008)}]{gadotti:pseudobulge.msigma}
{Gadotti}, D.~A., \& {Kauffmann}, G. 2008, \mnras, in press, arXiv:0811.1219
  [astro-ph]

\bibitem[{{Gallagher} \& {Ostriker}(1972)}]{gallagherostriker72}
{Gallagher}, III, J.~S., \& {Ostriker}, J.~P. 1972, \aj, 77, 288

\bibitem[{{Gallazzi} {et~al.}(2006){Gallazzi}, {Charlot}, {Brinchmann}, \&
  {White}}]{gallazzi06:ages}
{Gallazzi}, A., {Charlot}, S., {Brinchmann}, J., \& {White}, S.~D.~M. 2006,
  \mnras, 370, 1106

\bibitem[{{Genzel} {et~al.}(2008)}]{genzel:highz.rapid.secular}
{Genzel}, R., {et~al.} 2008, \apj, 687, 59

\bibitem[{{Gnedin} {et~al.}(2004){Gnedin}, {Kravtsov}, {Klypin}, \&
  {Nagai}}]{gnedin:adiabatic.contraction}
{Gnedin}, O.~Y., {Kravtsov}, A.~V., {Klypin}, A.~A., \& {Nagai}, D. 2004, \apj,
  616, 16

\bibitem[{{Graves} {et~al.}(2009){Graves}, {Faber}, \&
  {Schiavon}}]{graves:ssps.vs.fp.location}
{Graves}, G.~J., {Faber}, S.~M., \& {Schiavon}, R.~P. 2009, \apj, 698, 1590

\bibitem[{{Grazian} {et~al.}(2007)}]{grazian:drg.comparisons}
{Grazian}, A., {et~al.} 2007, \aap, 465, 393

\bibitem[{{Greene} {et~al.}(2008){Greene}, {Ho}, \&
  {Barth}}]{greene:pseudobulge.msigma}
{Greene}, J.~E., {Ho}, L.~C., \& {Barth}, A.~J. 2008, \apj, 688, 159

\bibitem[{{H{\"a}ring} \& {Rix}(2004)}]{haringrix}
{H{\"a}ring}, N., \& {Rix}, H.-W. 2004, \apjl, 604, L89

\bibitem[{{Hernquist} {et~al.}(1993){Hernquist}, {Spergel}, \&
  {Heyl}}]{hernquist:phasespace}
{Hernquist}, L., {Spergel}, D.~N., \& {Heyl}, J.~S. 1993, \apj, 416, 415

\bibitem[{{Hoffman} {et~al.}(2009){Hoffman}, {Cox}, {Dutta}, \&
  {Hernquist}}]{hoffman:dissipation.and.gal.kinematics}
{Hoffman}, L.~K., {Cox}, T.~J., {Dutta}, S.~N., \& {Hernquist}, L.~E. 2009,
  \apjl, in press, arXiv:0903.3064 [astro-ph]

\bibitem[{{Hopkins} \& {Beacom}(2006)}]{hopkinsbeacom:sfh}
{Hopkins}, A.~M., \& {Beacom}, J.~F. 2006, \apj, 651, 142

\bibitem[{{Hopkins} {et~al.}(2009{\natexlab{a}}){Hopkins}, {Bundy}, {Murray},
  {Quataert}, {Lauer}, \& {Ma}}]{hopkins:density.galcores}
{Hopkins}, P.~F., {Bundy}, K., {Murray}, N., {Quataert}, E., {Lauer}, T.~R., \&
  {Ma}, C.-P. 2009{\natexlab{a}}, \mnras, 398, 898

\bibitem[{{Hopkins} {et~al.}(2009{\natexlab{b}}){Hopkins}, {Cox}, {Dutta},
  {Hernquist}, {Kormendy}, \& {Lauer}}]{hopkins:cusps.ell}
{Hopkins}, P.~F., {Cox}, T.~J., {Dutta}, S.~N., {Hernquist}, L., {Kormendy},
  J., \& {Lauer}, T.~R. 2009{\natexlab{b}}, \apjs, 181, 135

\bibitem[{{Hopkins} {et~al.}(2008{\natexlab{a}}){Hopkins}, {Cox}, \&
  {Hernquist}}]{hopkins:cusps.fp}
{Hopkins}, P.~F., {Cox}, T.~J., \& {Hernquist}, L. 2008{\natexlab{a}}, \apj,
  689, 17

\bibitem[{{Hopkins} {et~al.}(2008{\natexlab{b}}){Hopkins}, {Cox}, {Kere{\v s}},
  \& {Hernquist}}]{hopkins:groups.ell}
{Hopkins}, P.~F., {Cox}, T.~J., {Kere{\v s}}, D., \& {Hernquist}, L.
  2008{\natexlab{b}}, \apjs, 175, 390

\bibitem[{{Hopkins} {et~al.}(2009{\natexlab{c}}){Hopkins}, {Cox}, {Younger}, \&
  {Hernquist}}]{hopkins:disk.survival}
{Hopkins}, P.~F., {Cox}, T.~J., {Younger}, J.~D., \& {Hernquist}, L.
  2009{\natexlab{c}}, \apj, 691, 1168

\bibitem[{{Hopkins} \& {Hernquist}(2006)}]{hopkins:seyferts}
{Hopkins}, P.~F., \& {Hernquist}, L. 2006, \apjs, 166, 1

\bibitem[{{Hopkins} {et~al.}(2005{\natexlab{a}}){Hopkins}, {Hernquist}, {Cox},
  {Di Matteo}, {Martini}, {Robertson}, \&
  {Springel}}]{hopkins:lifetimes.methods}
{Hopkins}, P.~F., {Hernquist}, L., {Cox}, T.~J., {Di Matteo}, T., {Martini},
  P., {Robertson}, B., \& {Springel}, V. 2005{\natexlab{a}}, \apj, 630, 705

\bibitem[{{Hopkins} {et~al.}(2008{\natexlab{c}}){Hopkins}, {Hernquist}, {Cox},
  {Dutta}, \& {Rothberg}}]{hopkins:cusps.mergers}
{Hopkins}, P.~F., {Hernquist}, L., {Cox}, T.~J., {Dutta}, S.~N., \& {Rothberg},
  B. 2008{\natexlab{c}}, \apj, 679, 156

\bibitem[{{Hopkins} {et~al.}(2008{\natexlab{d}}){Hopkins}, {Hernquist}, {Cox},
  \& {Kere{\v s}}}]{hopkins:groups.qso}
{Hopkins}, P.~F., {Hernquist}, L., {Cox}, T.~J., \& {Kere{\v s}}, D.
  2008{\natexlab{d}}, \apjs, 175, 356

\bibitem[{{Hopkins} {et~al.}(2009{\natexlab{d}}){Hopkins}, {Hernquist}, {Cox},
  {Kere{\v s}}, \& {Wuyts}}]{hopkins:cusps.evol}
{Hopkins}, P.~F., {Hernquist}, L., {Cox}, T.~J., {Kere{\v s}}, D., \& {Wuyts},
  S. 2009{\natexlab{d}}, \apj, 691, 1424

\bibitem[{{Hopkins} {et~al.}(2007{\natexlab{a}}){Hopkins}, {Hernquist}, {Cox},
  {Robertson}, \& {Krause}}]{hopkins:bhfp.theory}
{Hopkins}, P.~F., {Hernquist}, L., {Cox}, T.~J., {Robertson}, B., \& {Krause},
  E. 2007{\natexlab{a}}, \apj, 669, 45

\bibitem[{{Hopkins} {et~al.}(2007{\natexlab{b}}){Hopkins}, {Hernquist}, {Cox},
  {Robertson}, \& {Krause}}]{hopkins:bhfp.obs}
---. 2007{\natexlab{b}}, \apj, 669, 67

\bibitem[{{Hopkins} {et~al.}(2005{\natexlab{b}}){Hopkins}, {Hernquist},
  {Martini}, {Cox}, {Robertson}, {Di Matteo}, \&
  {Springel}}]{hopkins:lifetimes.letter}
{Hopkins}, P.~F., {Hernquist}, L., {Martini}, P., {Cox}, T.~J., {Robertson},
  B., {Di Matteo}, T., \& {Springel}, V. 2005{\natexlab{b}}, \apjl, 625, L71

\bibitem[{{Hopkins} {et~al.}(2009{\natexlab{e}}){Hopkins}, {Lauer}, {Cox},
  {Hernquist}, \& {Kormendy}}]{hopkins:cores}
{Hopkins}, P.~F., {Lauer}, T.~R., {Cox}, T.~J., {Hernquist}, L., \& {Kormendy},
  J. 2009{\natexlab{e}}, \apjs, 181, 486

\bibitem[{{Hopkins} {et~al.}(2009{\natexlab{f}}){Hopkins}, {Murray},
  {Quataert}, \& {Thompson}}]{hopkins:maximum.surface.densities}
{Hopkins}, P.~F., {Murray}, N., {Quataert}, E., \& {Thompson}, T.~A.
  2009{\natexlab{f}}, \mnras, in press, arXiv:0908.4088

\bibitem[{{Hopkins} {et~al.}(2007{\natexlab{c}}){Hopkins}, {Richards}, \&
  {Hernquist}}]{hopkins:bol.qlf}
{Hopkins}, P.~F., {Richards}, G.~T., \& {Hernquist}, L. 2007{\natexlab{c}},
  \apj, 654, 731

\bibitem[{{Hopkins} {et~al.}(2006){Hopkins}, {Robertson}, {Krause},
  {Hernquist}, \& {Cox}}]{hopkins:msigma.limit}
{Hopkins}, P.~F., {Robertson}, B., {Krause}, E., {Hernquist}, L., \& {Cox},
  T.~J. 2006, \apj, 652, 107

\bibitem[{{Hopkins} {et~al.}(2009{\natexlab{g}})}]{hopkins:merger.rates}
{Hopkins}, P.~F., {et~al.} 2009{\natexlab{g}}, \mnras, in press,
  arXiv:0906.5357

\bibitem[{{Hu}(2008)}]{hu:msigma.pseudobulges}
{Hu}, J. 2008, \mnras, 386, 2242

\bibitem[{{Hyde} \& {Bernardi}(2009)}]{hyde:scaling.relations.curvature}
{Hyde}, J.~B., \& {Bernardi}, M. 2009, \mnras, 349

\bibitem[{{Ilbert} {et~al.}(2009)}]{ilbert:cosmos.morph.mfs}
{Ilbert}, O., {et~al.} 2009, \apj, in press, arXiv:0903.0102

\bibitem[{{Jesseit} {et~al.}(2008){Jesseit}, {Cappellari}, {Naab}, {Emsellem},
  \& {Burkert}}]{jesseit:merger.rem.spin.vs.gas}
{Jesseit}, R., {Cappellari}, M., {Naab}, T., {Emsellem}, E., \& {Burkert}, A.
  2008, \mnras, in press, arXiv:0810.0137 [astro-ph]

\bibitem[{{Jesseit} {et~al.}(2007){Jesseit}, {Naab}, {Peletier}, \&
  {Burkert}}]{jesseit:kinematics}
{Jesseit}, R., {Naab}, T., {Peletier}, R.~F., \& {Burkert}, A. 2007, \mnras,
  376, 997

\bibitem[{{Jogee} {et~al.}(2005){Jogee}, {Scoville}, \&
  {Kenney}}]{jogee:H2.masses}
{Jogee}, S., {Scoville}, N., \& {Kenney}, J.~D.~P. 2005, \apj, 630, 837

\bibitem[{{Kannappan}(2004)}]{kannappan:gfs}
{Kannappan}, S.~J. 2004, \apjl, 611, L89

\bibitem[{{Kere{\v s}} {et~al.}(2005){Kere{\v s}}, {Katz}, {Weinberg}, \&
  {Dav{\'e}}}]{keres:hot.halos}
{Kere{\v s}}, D., {Katz}, N., {Weinberg}, D.~H., \& {Dav{\'e}}, R. 2005,
  \mnras, 363, 2

\bibitem[{{Khochfar} \& {Silk}(2006)}]{khochfar:size.evolution.model}
{Khochfar}, S., \& {Silk}, J. 2006, \apjl, 648, L21

\bibitem[{{Komatsu} {et~al.}(2009)}]{komatsu:wmap5}
{Komatsu}, E., {et~al.} 2009, \apjs, 180, 330

\bibitem[{{Kormendy}(1985)}]{kormendy:spheroidal1}
{Kormendy}, J. 1985, \apj, 295, 73

\bibitem[{{Kormendy} {et~al.}(2009){Kormendy}, {Fisher}, {Cornell}, \&
  {Bender}}]{jk:profiles}
{Kormendy}, J., {Fisher}, D.~B., {Cornell}, M.~E., \& {Bender}, R. 2009, \apjs,
  182, 216

\bibitem[{{Kriek} {et~al.}(2008{\natexlab{a}}){Kriek}, {van der Wel}, {van
  Dokkum}, {Franx}, \& {Illingworth}}]{kriek:highz.red.sequence}
{Kriek}, M., {van der Wel}, A., {van Dokkum}, P.~G., {Franx}, M., \&
  {Illingworth}, G.~D. 2008{\natexlab{a}}, \apj, 682, 896

\bibitem[{{Kriek} {et~al.}(2006)}]{kriek:drg.seds}
{Kriek}, M., {et~al.} 2006, \apjl, 649, L71

\bibitem[{{Kriek}
  {et~al.}(2008{\natexlab{b}})}]{kriek:08.nir.spectroscopy.highz}
---. 2008{\natexlab{b}}, \apj, 677, 219

\bibitem[{{Kuntschner} {et~al.}(2006)}]{kuntschner:line.strength.maps}
{Kuntschner}, H., {et~al.} 2006, \mnras, 369, 497

\bibitem[{{Labb{\'e}} {et~al.}(2005)}]{labbe05:drgs}
{Labb{\'e}}, I., {et~al.} 2005, \apjl, 624, L81

\bibitem[{{Lauer} {et~al.}(2007{\natexlab{a}}){Lauer}, {Tremaine}, {Richstone},
  \& {Faber}}]{lauer:mbh.bias}
{Lauer}, T.~R., {Tremaine}, S., {Richstone}, D., \& {Faber}, S.~M.
  2007{\natexlab{a}}, \apj, 670, 249

\bibitem[{{Lauer} {et~al.}(2007{\natexlab{b}})}]{lauer:bimodal.profiles}
{Lauer}, T.~R., {et~al.} 2007{\natexlab{b}}, \apj, 664, 226

\bibitem[{{Lauer} {et~al.}(2007{\natexlab{c}})}]{lauer:massive.bhs}
---. 2007{\natexlab{c}}, \apj, 662, 808

\bibitem[{{Lin} {et~al.}(2008)}]{lin:mergers.by.type}
{Lin}, L., {et~al.} 2008, \apj, 681, 232

\bibitem[{{Longhetti} {et~al.}(2007){Longhetti}, {Saracco}, {Severgnini},
  {Della Ceca}, {Mannucci}, {Bender}, {Drory}, {Feulner}, \&
  {Hopp}}]{longhetti:2007.z15.kormendy.relation}
{Longhetti}, M., {Saracco}, P., {Severgnini}, P., {Della Ceca}, R., {Mannucci},
  F., {Bender}, R., {Drory}, N., {Feulner}, G., \& {Hopp}, U. 2007, \mnras,
  374, 614

\bibitem[{{Maller} {et~al.}(2006){Maller}, {Katz}, {Kere{\v s}}, {Dav{\'e}}, \&
  {Weinberg}}]{maller:sph.merger.rates}
{Maller}, A.~H., {Katz}, N., {Kere{\v s}}, D., {Dav{\'e}}, R., \& {Weinberg},
  D.~H. 2006, \apj, 647, 763

\bibitem[{{Mannucci} {et~al.}(2009)}]{mannucci:z3.gal.gfs.tf}
{Mannucci}, F., {et~al.} 2009, \mnras, in press, arXiv:0902.2398

\bibitem[{{Maraston}(2005)}]{maraston:ssps}
{Maraston}, C. 2005, \mnras, 362, 799

\bibitem[{{Maraston} {et~al.}(2006){Maraston}, {Daddi}, {Renzini}, {Cimatti},
  {Dickinson}, {Papovich}, {Pasquali}, \& {Pirzkal}}]{maraston:ssp.effects}
{Maraston}, C., {Daddi}, E., {Renzini}, A., {Cimatti}, A., {Dickinson}, M.,
  {Papovich}, C., {Pasquali}, A., \& {Pirzkal}, N. 2006, \apj, 652, 85

\bibitem[{{Marchesini} {et~al.}(2008){Marchesini}, {van Dokkum}, {Forster
  Schreiber}, {Franx}, {Labbe'}, \& {Wuyts}}]{marchesini:highz.stellar.mfs}
{Marchesini}, D., {van Dokkum}, P.~G., {Forster Schreiber}, N.~M., {Franx}, M.,
  {Labbe'}, I., \& {Wuyts}, S. 2008, \apj, in press, arXiv:0811.1773 [astro-ph]

\bibitem[{{McDermid} {et~al.}(2006)}]{mcdermid:sauron.profiles}
{McDermid}, R.~M., {et~al.} 2006, \mnras, 373, 906

\bibitem[{{McGaugh}(2005)}]{mcgaugh:tf}
{McGaugh}, S.~S. 2005, \apj, 632, 859

\bibitem[{{McLure} \& {Dunlop}(2004)}]{mclure.dunlop:mbh}
{McLure}, R.~J., \& {Dunlop}, J.~S. 2004, \mnras, 352, 1390

\bibitem[{{Merloni} {et~al.}(2004){Merloni}, {Rudnick}, \& {Di
  Matteo}}]{merloni:magorrian.evolution}
{Merloni}, A., {Rudnick}, G., \& {Di Matteo}, T. 2004, \mnras, 354, L37

\bibitem[{{Mihos} \& {Hernquist}(1994{\natexlab{a}})}]{mihos:cusps}
{Mihos}, J.~C., \& {Hernquist}, L. 1994{\natexlab{a}}, \apjl, 437, L47

\bibitem[{{Mihos} \& {Hernquist}(1994{\natexlab{b}})}]{mihos:starbursts.94}
---. 1994{\natexlab{b}}, \apjl, 431, L9

\bibitem[{{Mihos} \& {Hernquist}(1996)}]{mihos:starbursts.96}
---. 1996, \apj, 464, 641

\bibitem[{{Milosavljevi{\'c}} {et~al.}(2002){Milosavljevi{\'c}}, {Merritt},
  {Rest}, \& {van den Bosch}}]{milosavljevic:core.mass}
{Milosavljevi{\'c}}, M., {Merritt}, D., {Rest}, A., \& {van den Bosch}, F.~C.
  2002, \mnras, 331, L51

\bibitem[{{Naab} {et~al.}(2006{\natexlab{a}}){Naab}, {Jesseit}, \&
  {Burkert}}]{naab:gas}
{Naab}, T., {Jesseit}, R., \& {Burkert}, A. 2006{\natexlab{a}}, \mnras, 372,
  839

\bibitem[{{Naab} {et~al.}(2009){Naab}, {Johansson}, \&
  {Ostriker}}]{naab:size.evol.from.minor.mergers}
{Naab}, T., {Johansson}, P.~H., \& {Ostriker}, J.~P. 2009, \apj, in press,
  arXiv:0903.1636

\bibitem[{{Naab} {et~al.}(2006{\natexlab{b}}){Naab}, {Khochfar}, \&
  {Burkert}}]{naab:dry.mergers}
{Naab}, T., {Khochfar}, S., \& {Burkert}, A. 2006{\natexlab{b}}, \apjl, 636,
  L81

\bibitem[{{Naab} \& {Trujillo}(2006)}]{naab:profiles}
{Naab}, T., \& {Trujillo}, I. 2006, \mnras, 369, 625

\bibitem[{{O{\~n}orbe} {et~al.}(2006){O{\~n}orbe}, {Dom{\'{\i}}nguez-Tenreiro},
  {S{\'a}iz}, {Artal}, \& {Serna}}]{onorbe:diss.fp.details}
{O{\~n}orbe}, J., {Dom{\'{\i}}nguez-Tenreiro}, R., {S{\'a}iz}, A., {Artal}, H.,
  \& {Serna}, A. 2006, \mnras, 373, 503

\bibitem[{{Pannella} {et~al.}(2006){Pannella}, {Hopp}, {Saglia}, {Bender},
  {Drory}, {Salvato}, {Gabasch}, \& {Feulner}}]{pannella:mfs}
{Pannella}, M., {Hopp}, U., {Saglia}, R.~P., {Bender}, R., {Drory}, N.,
  {Salvato}, M., {Gabasch}, A., \& {Feulner}, G. 2006, \apjl, 639, L1

\bibitem[{{Peletier} {et~al.}(1990){Peletier}, {Davies}, {Illingworth},
  {Davis}, \& {Cawson}}]{peletier:profiles}
{Peletier}, R.~F., {Davies}, R.~L., {Illingworth}, G.~D., {Davis}, L.~E., \&
  {Cawson}, M. 1990, \aj, 100, 1091

\bibitem[{{Peng} {et~al.}(2006){Peng}, {Impey}, {Rix}, {Kochanek}, {Keeton},
  {Falco}, {Leh{\'a}r}, \& {McLeod}}]{peng:magorrian.evolution}
{Peng}, C.~Y., {Impey}, C.~D., {Rix}, H.-W., {Kochanek}, C.~S., {Keeton},
  C.~R., {Falco}, E.~E., {Leh{\'a}r}, J., \& {McLeod}, B.~A. 2006, \apj, 649,
  616

\bibitem[{{P{\'e}rez-Gonz{\'a}lez}
  {et~al.}(2008)}]{perezgonzalez:mf.compilation}
{P{\'e}rez-Gonz{\'a}lez}, P.~G., {et~al.} 2008, \apj, 675, 234

\bibitem[{{Pozzetti} {et~al.}(2009)}]{pozzetti:2009.z1.gal.bimodal.mf}
{Pozzetti}, L., {et~al.} 2009, \aap, in press, arXiv:0907.5416

\bibitem[{{Puech} {et~al.}(2008)}]{puech:tf.evol}
{Puech}, M., {et~al.} 2008, \aap, 484, 173

\bibitem[{{Ravindranath} {et~al.}(2004)}]{ravindranath:disk.size.evol}
{Ravindranath}, S., {et~al.} 2004, \apjl, 604, L9

\bibitem[{{Rettura} {et~al.}(2008){Rettura}, {Rosati}, {Nonino}, {Fosbury},
  {Gobat}, {Menci}, {Strazzullo}, {Mei}, {Demarco}, \&
  {Ford}}]{rettura:2008.z1.sizes.clustervsfield}
{Rettura}, A., {Rosati}, P., {Nonino}, M., {Fosbury}, R.~A.~E., {Gobat}, R.,
  {Menci}, N., {Strazzullo}, V., {Mei}, S., {Demarco}, R., \& {Ford}, H.~C.
  2008, \apj, in press, arXiv:0806.4604

\bibitem[{{Robertson} {et~al.}(2006{\natexlab{a}}){Robertson}, {Cox},
  {Hernquist}, {Franx}, {Hopkins}, {Martini}, \& {Springel}}]{robertson:fp}
{Robertson}, B., {Cox}, T.~J., {Hernquist}, L., {Franx}, M., {Hopkins}, P.~F.,
  {Martini}, P., \& {Springel}, V. 2006{\natexlab{a}}, \apj, 641, 21

\bibitem[{{Robertson} {et~al.}(2006{\natexlab{b}}){Robertson}, {Hernquist},
  {Cox}, {Di Matteo}, {Hopkins}, {Martini}, \&
  {Springel}}]{robertson:msigma.evolution}
{Robertson}, B., {Hernquist}, L., {Cox}, T.~J., {Di Matteo}, T., {Hopkins},
  P.~F., {Martini}, P., \& {Springel}, V. 2006{\natexlab{b}}, \apj, 641, 90

\bibitem[{{Rothberg} \& {Joseph}(2004)}]{rj:profiles}
{Rothberg}, B., \& {Joseph}, R.~D. 2004, \aj, 128, 2098

\bibitem[{{Salviander} {et~al.}(2006){Salviander}, {Shields}, {Gebhardt}, \&
  {Bonning}}]{salviander:msigma.evolution}
{Salviander}, S., {Shields}, G.~A., {Gebhardt}, K., \& {Bonning}, E.~W. 2006,
  New Astronomy Review, 50, 803

\bibitem[{{Salviander} {et~al.}(2007){Salviander}, {Shields}, {Gebhardt}, \&
  {Bonning}}]{salviander:midz.msigma.evol}
---. 2007, \apj, 662, 131

\bibitem[{{S{\'a}nchez-Bl{\'a}zquez} {et~al.}(2007){S{\'a}nchez-Bl{\'a}zquez},
  {Forbes}, {Strader}, {Brodie}, \& {Proctor}}]{sanchezblazquez:ssp.gradients}
{S{\'a}nchez-Bl{\'a}zquez}, P., {Forbes}, D.~A., {Strader}, J., {Brodie}, J.,
  \& {Proctor}, R. 2007, \mnras, 377, 759

\bibitem[{{Sersic}(1968)}]{sersic:profile}
{Sersic}, J.~L. 1968, {Atlas de galaxias australes} (Cordoba, Argentina:
  Observatorio Astronomico, 1968)

\bibitem[{{Shapley} {et~al.}(2005){Shapley}, {Coil}, {Ma}, \&
  {Bundy}}]{shapley:z1.abundances}
{Shapley}, A.~E., {Coil}, A.~L., {Ma}, C.-P., \& {Bundy}, K. 2005, \apj, 635,
  1006

\bibitem[{{Shen} {et~al.}(2003){Shen}, {Mo}, {White}, {Blanton}, {Kauffmann},
  {Voges}, {Brinkmann}, \& {Csabai}}]{shen:size.mass}
{Shen}, S., {Mo}, H.~J., {White}, S.~D.~M., {Blanton}, M.~R., {Kauffmann}, G.,
  {Voges}, W., {Brinkmann}, J., \& {Csabai}, I. 2003, \mnras, 343, 978

\bibitem[{{Silk} \& {Rees}(1998)}]{silkrees:msigma}
{Silk}, J., \& {Rees}, M.~J. 1998, \aap, 331, L1

\bibitem[{{Somerville} {et~al.}(2008)}]{somerville:disk.size.evol}
{Somerville}, R.~S., {et~al.} 2008, \apj, 672, 776

\bibitem[{{Springel} {et~al.}(2005{\natexlab{a}}){Springel}, {Di Matteo}, \&
  {Hernquist}}]{springel:red.galaxies}
{Springel}, V., {Di Matteo}, T., \& {Hernquist}, L. 2005{\natexlab{a}}, \apjl,
  620, L79

\bibitem[{{Springel} {et~al.}(2005{\natexlab{b}}){Springel}, {Di Matteo}, \&
  {Hernquist}}]{springel:models}
---. 2005{\natexlab{b}}, \mnras, 361, 776

\bibitem[{{Springel} \& {Hernquist}(2005)}]{springel:spiral.in.merger}
{Springel}, V., \& {Hernquist}, L. 2005, \apjl, 622, L9

\bibitem[{{Tacconi} {et~al.}(2002){Tacconi}, {Genzel}, {Lutz}, {Rigopoulou},
  {Baker}, {Iserlohe}, \& {Tecza}}]{tacconi:ulirgs.sb.profiles}
{Tacconi}, L.~J., {Genzel}, R., {Lutz}, D., {Rigopoulou}, D., {Baker}, A.~J.,
  {Iserlohe}, C., \& {Tecza}, M. 2002, \apj, 580, 73

\bibitem[{{Tacconi} {et~al.}(2008)}]{tacconi:smg.mgr.lifetime.to.quiescent}
{Tacconi}, L.~J., {et~al.} 2008, \apj, 680, 246

\bibitem[{{Taylor} {et~al.}(2009){Taylor}, {Franx}, {Glazebrook}, {Brinchmann},
  {van der Wel}, \& {van Dokkum}}]{taylor:2009.no.compact.massive.local.gal}
{Taylor}, E.~N., {Franx}, M., {Glazebrook}, K., {Brinchmann}, J., {van der
  Wel}, A., \& {van Dokkum}, P.~G. 2009, \apj, in press, arXiv:0907.4766

\bibitem[{{Toft} {et~al.}(2007)}]{toft:z2.sizes.vs.sfr}
{Toft}, S., {et~al.} 2007, \apj, 671, 285

\bibitem[{{Tremaine} {et~al.}(2002)}]{tremaine:msigma}
{Tremaine}, S., {et~al.} 2002, \apj, 574, 740

\bibitem[{{Treu} {et~al.}(2007){Treu}, {Woo}, {Malkan}, \&
  {Blandford}}]{treu:msigma.evol}
{Treu}, T., {Woo}, J.-H., {Malkan}, M.~A., \& {Blandford}, R.~D. 2007, \apj,
  667, 117

\bibitem[{{Trujillo} {et~al.}(2009){Trujillo}, {Cenarro}, {de
  Lorenzo-C{\'a}ceres}, {Vazdekis}, {de la Rosa}, \&
  {Cava}}]{trujillo:dense.gal.nearby}
{Trujillo}, I., {Cenarro}, A.~J., {de Lorenzo-C{\'a}ceres}, A., {Vazdekis}, A.,
  {de la Rosa}, I.~G., \& {Cava}, A. 2009, \apjl, 692, L118

\bibitem[{{Trujillo} {et~al.}(2007){Trujillo}, {Conselice}, {Bundy}, {Cooper},
  {Eisenhardt}, \& {Ellis}}]{trujillo:ell.size.evol.update}
{Trujillo}, I., {Conselice}, C.~J., {Bundy}, K., {Cooper}, M.~C., {Eisenhardt},
  P., \& {Ellis}, R.~S. 2007, \mnras, 382, 109

\bibitem[{{Trujillo} {et~al.}(2004)}]{trujillo:size.mass.to.z3}
{Trujillo}, I., {et~al.} 2004, \apj, 604, 521

\bibitem[{{Trujillo}
  {et~al.}(2006{\natexlab{a}})}]{trujillo:compact.most.massive}
---. 2006{\natexlab{a}}, \mnras, 373, L36

\bibitem[{{Trujillo} {et~al.}(2006{\natexlab{b}})}]{trujillo:size.evolution}
---. 2006{\natexlab{b}}, \apj, 650, 18

\bibitem[{{Valentinuzzi} {et~al.}(2009){Valentinuzzi}, {Fritz}, {Poggianti},
  {Bettoni}, {Cava}, {Fasano}, {D'Onofrio}, {Couch}, {Dressler}, {Moles},
  {Moretti}, {Omizzolo}, {Kjaergaard}, {Vanzella}, \&
  {Varela}}]{valentinuzzi:superdense.local.ell.wings}
{Valentinuzzi}, T., {Fritz}, J., {Poggianti}, B.~M., {Bettoni}, D., {Cava}, A.,
  {Fasano}, G., {D'Onofrio}, M., {Couch}, W.~J., {Dressler}, A., {Moles}, M.,
  {Moretti}, A., {Omizzolo}, A., {Kjaergaard}, P., {Vanzella}, E., \& {Varela},
  J. 2009, \apj, in press, arXiv:0907.2392

\bibitem[{{van der Wel} {et~al.}(2009){van der Wel}, {Bell}, {van den Bosch},
  {Gallazzi}, \& {Rix}}]{vanderwel:size.numden.massive.ell}
{van der Wel}, A., {Bell}, E.~F., {van den Bosch}, F.~C., {Gallazzi}, A., \&
  {Rix}, H.-W. 2009, \apj, 698, 1232

\bibitem[{{van der Wel} {et~al.}(2008){van der Wel}, {Holden}, {Zirm}, {Franx},
  {Rettura}, {Illingworth}, \& {Ford}}]{vanderwel:z1.compact.ell}
{van der Wel}, A., {Holden}, B.~P., {Zirm}, A.~W., {Franx}, M., {Rettura}, A.,
  {Illingworth}, G.~D., \& {Ford}, H.~C. 2008, \apj, 688, 48

\bibitem[{{van der Wel} \& {van der Marel}(2008)}]{vanderwel:ell.vsigma.evol}
{van der Wel}, A., \& {van der Marel}, R.~P. 2008, \apj, 684, 260

\bibitem[{{van Dokkum} {et~al.}(2008)}]{vandokkum:z2.sizes}
{van Dokkum}, P., {et~al.} 2008, \apjl, 677, L5

\bibitem[{{van Dokkum}(2005)}]{vandokkum:dry.mergers}
{van Dokkum}, P.~G. 2005, \aj, 130, 2647

\bibitem[{{van Dokkum}(2008)}]{vandokkum:imf.evol}
---. 2008, \apj, 674, 29

\bibitem[{{van Dokkum} {et~al.}(2009){van Dokkum}, {Kriek}, \&
  {Franx}}]{vandokkum:high.sigma.compact.gal}
{van Dokkum}, P.~G., {Kriek}, M., \& {Franx}, M. 2009, \nat, in press
  [arXiv:0906.2778]

\bibitem[{{van Dokkum} {et~al.}(2006)}]{vandokkum06:drgs}
{van Dokkum}, P.~G., {et~al.} 2006, \apjl, 638, L59

\bibitem[{{von der Linden} {et~al.}(2007){von der Linden}, {Best}, {Kauffmann},
  \& {White}}]{vonderlinden:bcg.scaling.relations}
{von der Linden}, A., {Best}, P.~N., {Kauffmann}, G., \& {White}, S.~D.~M.
  2007, \mnras, 379, 867

\bibitem[{{Woo} {et~al.}(2006){Woo}, {Treu}, {Malkan}, \&
  {Blandford}}]{woo06:lowz.msigma.evolution}
{Woo}, J.-H., {Treu}, T., {Malkan}, M.~A., \& {Blandford}, R.~D. 2006, \apj,
  645, 900

\bibitem[{{Wuyts} {et~al.}(2009){Wuyts}, {Franx}, {Cox}, {Hernquist},
  {Hopkins}, {Robertson}, \& {van Dokkum}}]{wuyts:photometry.biases.mgrrem}
{Wuyts}, S., {Franx}, M., {Cox}, T.~J., {Hernquist}, L., {Hopkins}, P.~F.,
  {Robertson}, B.~E., \& {van Dokkum}, P.~G. 2009, \apj, 696, 348

\bibitem[{{Wuyts} {et~al.}(2007)}]{wuyts:irac.drg.colors}
{Wuyts}, S., {et~al.} 2007, \apj, 655, 51

\bibitem[{{Younger} {et~al.}(2008{\natexlab{a}}){Younger}, {Hopkins}, {Cox}, \&
  {Hernquist}}]{younger:minor.mergers}
{Younger}, J.~D., {Hopkins}, P.~F., {Cox}, T.~J., \& {Hernquist}, L.
  2008{\natexlab{a}}, \apj, 686, 815

\bibitem[{{Younger} {et~al.}(2008{\natexlab{b}})}]{younger:smg.sizes}
{Younger}, J.~D., {et~al.} 2008{\natexlab{b}}, \apj, 688, 59

\bibitem[{{Zhao}(2002)}]{zhao:2002.adiabatic.expansion.stellarwinds}
{Zhao}, H. 2002, \mnras, 336, 159

\bibitem[{{Zirm} {et~al.}(2007)}]{zirm:drg.sizes}
{Zirm}, A.~W., {et~al.} 2007, \apj, 656, 66

\end{thebibliography}

\end{document}